\documentclass[12pt]{article}
\newcommand{\ra}[1]{\renewcommand{\arraystretch}{#1}}

\usepackage{arxiv}
 
\usepackage{setspace}
\usepackage[table]{xcolor}    % loads also »colortbl«

\usepackage[utf8]{inputenc} % allow utf-8 input
\usepackage[T1]{fontenc}    % use 8-bit T1 fonts
\usepackage{hyperref}       % hyperlinks
\usepackage{url}            % simple URL typesetting
\usepackage{booktabs}       % professional-quality tables
\usepackage{amsfonts}       % blackboard math symbols
\usepackage{nicefrac}       % compact symbols for 1/2, etc.
\usepackage{microtype}      % microtypography
\usepackage{lipsum}

\usepackage{caption}
\usepackage{graphicx}
\usepackage{lscape}
\usepackage{hyperref}
\usepackage{amsmath}
\usepackage{longtable}%
\usepackage{colortbl}%

\usepackage{todonotes}

\renewcommand{\arraystretch}{1}

\title{The Geography of Open Source Software: Evidence from GitHub}

\author{
  Johannes Wachs\\
  Vienna Univ. of Economics and Business\\
  Complexity Science Hub Vienna\\
  \texttt{johannes.wachs@wu.ac.at} \\
   \And
 Mariusz Nitecki \\
  Vienna Univ. of Economics and Business\\
    \And
 William Schueller \\
  Medical Univ. of Vienna\\
  Complexity Science Hub Vienna
   \And
Axel Polleres \\
  Vienna Univ. of Economics and Business\\
  Complexity Science Hub Vienna\\
}

\begin{document}
\maketitle
\date{}
\begin{abstract}

Open Source Software (OSS) plays an important role in the digital economy. Yet although software production is amenable to remote collaboration and its outputs are easily shared across distances, software development seems to cluster geographically in places such as Silicon Valley, London, or Berlin. And while recent work indicates that OSS activity creates positive externalities which accrue locally through knowledge spillovers and information effects, up-to-date data on the geographic distribution of active open source developers is limited. This presents a significant blindspot for policymakers, who tend to promote OSS at the national level as a cost-saving tool for public sector institutions. We address this gap by geolocating more than half a million active contributors to GitHub in early 2021 at various spatial scales. Compared to results from 2010, we find a significant increase in the share of developers based in Asia, Latin America and Eastern Europe, suggesting a more even spread of OSS developers globally. Within countries, however, we find significant concentration in regions, exceeding the concentration of workers in high-tech fields. Social and economic development indicators predict at most half of regional variation in OSS activity in the EU, suggesting that clusters of OSS have idiosyncratic roots. We argue that policymakers seeking to foster OSS should focus locally rather than nationally, using the tools of cluster policy to support networks of OSS developers.

\keywords{Geography, open source software, GitHub, innovation, cluster policy}
\end{abstract}

\section*{Introduction}
The importance of software, both as a ubiquitous complement to other activities in the modern economy and as a key sector in its own right, is widely acknowledged \cite{andreessen2011software,nagle2019open}. Software is also an especially global industry, in part because its end products are easily shared and distributed through the Web. Yet the notion that the software industry might transcend geographic constraints is inconsistent with anecdotal observations that its most coveted jobs and leading firms tend to cluster in particular places like Silicon Valley, London, or Berlin \cite{grier2015tyranny}. These places have an out-sized influence on the art and practice of software engineering \cite{takhteyev2012coding}, which is increasingly shaping our economy and society \cite{wagner2021algo}. They also benefit from significant positive local externalities \cite{nagle2020report}. Despite these observations, researchers and policymakers have limited information about the extent of spatial clustering of software development and the socioeconomic conditions in these hotspots.

We address this gap by studying the geography of open source software (OSS) developers. Within the vast ecosystem of software, OSS plays a distinguished role \cite{eghbal2020working}, and is sometimes described as the infrastructure of the digital society. One recent estimate is that OSS contributes about 60 to 95 billion Euros annually to Eurozone GDP, and that a 10\% increase in OSS activity would lead to roughly 600 additional ICT startups in the EU  \cite{blind2021impact}. Mirroring trends in closed-source software, OSS contributions are thought to be intensely concentrated in space, despite the low cost of distributing software, the development of technologies for remote collaboration, and its inherently open nature. Research from 2010 estimated that over 7\% of global OSS activity at the time took place in or around San Francisco \cite{takhteyev2010investigating}. As OSS activity is known to impact local firm productivity \cite{nagle2018learning,nagle2019open} and rates of technology entrepreneurship \cite{wright2020open}, its geographic clustering likely effects economic growth and inequality, hence should be of interest to policymakers \cite{nagle2019government}. Policy interventions to date have sought to foster OSS for example through procurement regulations or trade and industrial policy \cite{blind2021impact}. These policies are usually motivated by the potential cost-savings of OSS adoption by public sector institutions, instead of the positive externalities that OSS production creates for local economies.

Geographic disparities may also influence who participates in OSS, and understanding them better can inform us about root causes of diversity issues in software \cite{albusays2021diversity,may2019gender,prana2021including} and its role in innovation \cite{dahlander2021open}. Though OSS is thought to be a highly decentralized activity in the global economy, previous work has shown how core places in the software industry have incredible influence over the way software is written around the world \cite{takhteyev2012coding}, for example by setting the norms and practices of software development used around the world. And though some breakthrough OSS innovations come from places such as Japan (Ruby), Finland (Linux), and South Africa (Ubuntu), they are often refined and popularized through central places like Chicago (where Ruby on Rails was created), the Pacific Northwest (where the creator of Linux now lives and works), and London (where Ubuntu is maintained) \cite{takhteyev2010investigating}.

So while the geographic distribution of OSS contributors likely has important localized effects on our society and economy, we lack an up-to-date geographic mapping of OSS activity. Previous work was either carried out over ten years ago \cite{gonzalez2008geographic,takhteyev2010investigating}, focuses on data at the country-level \cite{nagle2019government,wright2020open}, and/or on a subset of projects \cite{prana2021including}. We suggest that there is need for a fresh look at the geographic distribution of OSS developers, including regional data. This gap is becoming more relevant as researchers take a greater interest in the influence of algorithms and software on society \cite{wagner2021algo}. If algorithms do shape the fabric of our society, the hypothesis that software is created by a distinct group of people living in a few particular places merits testing. At the same time, software engineering researchers have gathered massive datasets on OS contributions \cite{gousios2012ghtorrent,pietri2019software,fry2020dataset}, presenting opportunities to study this question in greater depth.

In this work we implement a pipeline to geolocate highly active OSS developers on GitHub, cleaning and sharing the resulting data at both national and subnational levels. In contrast to studies from over ten years ago, we find a slightly more even global distribution of OSS activity, with marked growth among Asian, Latin American, and Eastern European countries. Within countries, however, we find that OSS remains highly concentrated in particular regions. OSS activity is even more spatially concentrated than the university educated population and workers in high-tech sectors. To better understand the structural conditions of geographic OSS hubs, we using a regression framework to relate OSS activity with social, political, and economic features of countries. No one feature can predict most of the variance observed between countries. A spatial econometric study of the EU regions replicates this finding at a more granular scale. This suggests that there are many necessary but not sufficient conditions for OSS clusters, with important implications for policymakers seeking to encourage OSS activity.

The rest of the paper is organized as follows. We first discuss related works on the importance of OSS in the digital economy both in general and locally. We then introduce our data collection process, describe how we geolocate developers, and share access to data on counts of active developers in various geographic units. We then analyze the distribution of developers at both the country and regional levels. We conclude with a discussion of limitations, potential uses for this dataset, and ideas for future work in this area. 

\section*{Literature Review}
In this section we review the literature on why industries cluster geographically in general, and reflect on ways in which the software industry and OSS in particular are unique in this context. We then survey emerging evidence on the local impact of OSS developers. Finally, we review the motivations for and results of policy interventions involving OSS. Research gaps between these concepts motivate our subsequent mapping of OSS activity and our regression analysis of correlated socio-economic factors.

\subsection*{Geographic Clusters and the case of Open Source Software}
It has long been observed that highly specialized activity at the frontier of knowledge clusters geographically \cite{krugman1991increasing}. Skilled people living and working near to one another interact and come up with new ideas, attracting more people with their own ideas \cite{watney2020clusters}. Network, information and agglomeration effects give these places comparative advantages that make up for pathologies of boom towns such as high rents. Many examples have been studied in the literature, for example bio-tech clusters in the US and China \cite{su2009spontaneous}, car manufacturing in Detroit \cite{klepper2010origin}, and the shoe industry in southern Italy \cite{boschma2007knowledge}. The latter study in particular highlights that co-location is insufficient to create the virtuous effects associated with successful clusters that are thought to accelerate innovation, entrepreneurship, and productivity growth. In particular it is the social networks and cognitive proximities of co-located firms and people of a cluster that make it sink or swim \cite{frenken2015industrial,juhasz2018creation,juhasz2021spinoffs}. Silicon Valley itself bounced back from significant setbacks in the 1980s due to its regional networks of people and institutions \cite{saxenian1990regional}. Despite the highly specific and seemingly irreplicable nature of individual clusters, they appear in a large variety of sectors.

The software industry is no exception. Indeed, even though software is easily shared and distributed, software creation has long been intensely geographically clustered \cite{bettencourt2002client}. A 2015 study claimed that 40\% of all jobs in the US software industry were clustered in merely nine cities \cite{grier2015tyranny} and Silicon Valley is a household name. As in other sectors, geographic proximity to customers, end-users, and other software developers improves developer productivity and software quality via spillover effects \cite{weterings2006impact}. Knowledge transfers between developers in the software industry often occur via informal networks, which complement more formal structures of collaboration and interaction \cite{trippl2009knowledge}.  

Within the world of software, OSS activity is also thought to be highly localized, even if long distance collaborations are frequent and frictionless thanks to the sophisticated collaboration tools OSS developers use. A study of activity in the 2000s estimated that roughly one in ten contributions to OSS libraries on GitHub \textit{globally} originated in the San Francisco Bay Area \cite{takhteyev2010investigating}. Several studies of OSS \cite{takhteyev2010investigating,lima2014coding,fackler2020gravity} have shown that the likelihood of collaboration decreases exponentially with distance.

In some sense it is remarkable that geography remains so important in OSS given that several of the mechanisms that explain why firms cluster do not clearly transfer to this specific context. For example, users and competitors in the OSS space are likely more geographically dispersed than in traditional industries. It seems that other mechanisms behind cluster strength and persistence, for instance network effects and knowledge spillovers, are enough to keep activity clustered even within the virtual, global collaboration network of OSS developers \cite{fershtman2011direct}. Contributors often only work on OSS in their free-time, next to a primary occupation, often in the software industry. In this way, the geographic concentration of OSS extends the concentration present in the software industry. The continued dominance of specific clusters, once they emerge, may be less of a puzzle: developers working in such places have significant advantages over their counterparts in peripheral regions. Indeed norms and practices of the software industry tend to flow from core to periphery locations \cite{takhteyev2012coding}.

\subsection*{Local Effects of OSS Development}
We have suggested that OSS development, like any other frontier knowledge creation, occurs in geographic clusters. At the same time OSS activity is generally unpaid and its outputs are public goods, distributed world wide via the web. And though OSS contributes substantial amounts to global GDP and productivity in this way \cite{greenstein2014digital}, it begs the question whether locations hosting OSS clusters benefit in the same way as they might from hosting, for example, a cluster of highly profitable biotech companies. While OSS activity does not generate taxable profits, recent work has shown that it generates significant benefits that accrue locally. Specifically, the public nature of OSS activity fosters entrepreneurship and investment by providing strong signals of quality. It also creates opportunities for developers and firms to learn by doing and to better integrate user feedback.

Recent work has shown that OSS activity is a strong predictor of IT entrepreneurship \cite{wright2020open}. Because of its open nature, OSS developers can effectively signal their specific areas of expertise, broadening the network of potential collaborators in a new venture. Public and transparent information about how people and teams work together makes it easier to attract outside investments \cite{kaminski2019new}, to get quality feedback \cite{wachs2021crowdfunding}, and to learn by observing \cite{riedl2018learning}. A large OSS footprint also suggests that a location has complementary assets necessary for software entrepreneurship, which can contribute to network effects. This presents OSS activity as an effective proxy measurement for a place's technological development and capacity for IT innovation. 

Beyond signaling effects, participation in OSS also brings many benefits. Firms contributing to OSS capture more value from the resulting software than free-riding competitors because they learn from the experience and gain valuable feedback \cite{nagle2018learning}. In general OSS is high quality because it benefits from the attention of a large, relatively independent crowd of users\cite{raymond1999cathedral,lakhani2004open,aksulu2010comprehensive}. Firms who use OSS benefit from this ecosystem \cite{nagle2019open}. Indeed these factors have led to growing adoption of and participation in OSS, including by firms with a traditionally proprietary orientation such as Microsoft and Apple \cite{anthes2016open}. Companies increasingly release in-house projects as OSS, for example Google's TensorFlow library for machine learning and Facebook's React web framework. Though companies cannot prevent competitors from using work that their employees contribute to OSS projects, they encourage such contributions anyway to gain legitimacy and access to communities \cite{dahlander2006man}. OSS development can also serve as an incubator for new software ventures, recalling the ``doing-using-interacting'' model of innovation \cite{jensen2007forms,alhusen2021new}. These virtuous effects of OSS activity manifest locally, through productivity growth in firms and new ventures.

Besides the specific outputs of the OSS sector and their direct uses, software know-how itself is clearly an important input in emerging sectors including Industry 4.0 \cite{balland2021mapping} and artificial intelligence, for instance via the intensely computational methods of deep-learning \cite{klinger2021deep}. It also contributes to productivity gains in manufacturing \cite{branstetter2019get}, suggesting that software has a complementary role in many sectors of the modern economy \cite{neffke2019value}, including for example biotechnology via bioinformatics \cite{hu2021bioinformatics}. In this way local activity in OSS, which is more easily observable than closed source software development, can serve as a valuable signal of the potential of a city, region, or country in the digital economy.

\subsection*{Public Policy and OSS}

Given its economic impact, it is not surprising that policymakers are increasingly interested in promoting OSS activity. These efforts often draw on insights from the cluster policy literature on traditional industries, but it is unclear if these ideas transfer to the OSS context. Many OSS contributions come from independent individuals, working in their spare time. OSS contributors only rarely receive direct financial support for their work, for instance through crowdfunding \cite{overney2020not} or consulting \cite{eghbal2020working}. Indeed they tend to be motivated by social and reputational reasons, and by specific technical problems \cite{gerosa2021shifting}. Economic gains from OSS activity tend to accrue only in the long run \cite{lerner2002some} and so are rarely counted on by contributors. So while cluster policy mainstays like SME tax incentives, networking events, and support for fundraising and branding \cite{uyarra2016impact} may help crystallize the economic potential of an active OSS scene, they are unlikely to attract or create new OSS contributors. In this way, optimal cluster policy for OSS activity merits additional investigation.

On the other hand, there has been significant effort by governments to encourage OSS activity in general. Lee \cite{lee2006government} and Blind \cite{blind2021impact} group motivations for public sector support of OSS into four groups: economic, technological, legal, and political. For example, using OSS saves on direct software costs and avoid vendor lock-in. Technologically, OSS solutions may be superior to closed source alternatives. Legally, governments, especially in the developing world, can avoid issues around software piracy by using OSS solutions. Finally, OSS use has political reasons: it increases transparency and eases public access to government IT infrastructure. These direct motivations and others have led to a significant amount of policy support OSS use.

An example of a successful policy promoting the use of OSS is the French law Circulaire 5608, passed in 2012 \cite{blind2021impact}. Motivated by potential cost-savings of avoiding closed source software licensing fees, the law requires French public bodies to consider open source solutions when procuring software. A study of the impact of this law by Nagle found significant social welfare gains \cite{nagle2019government} on the order of tens of millions of Euros. More significantly, second order effects observed include an increase in IT start-ups (9-18\% yearly increase vs. the counterfactual) and employment in IT employment (7-14\%), and a decrease in IT patents (5-16\%). In general, policy promoting OSS has focused on government purchasing and public sector use of software while less has been done to incentivize private sector or individual contributions to OSS \cite{blind2021impact}, let alone in terms of cluster policy. 

So while governments tend to have policy on OSS, it tends to focus on how the public sector can cut costs and improve transparency by using OSS, rather than how OSS adoption and development can improve local productivity and innovation outcomes. Public policy on the use of OSS also tends to happen at the national level, even though we will show that activity tends to cluster in regions. The rest of this work presents data on the actual geographic distribution of OSS developers and the characteristics of those places, both national and regional, which have many developers. Reflecting on these geographic distributions, we will revisit the question of effective policies for the promotion of OSS development in our conclusion.

\section*{Research Design and Data}
We now describe our data collection and processing pipeline. Before gathering information about activity levels and positions of individual OSS contributors, one needs to define contributors and their activity. Developers share code using version control protocols -- allowing them to track changes, compare, test and merge with modifications of others -- on specific online platforms. The most widely used protocol is called \textit{git}, and the most widely used public platform for projects using git is \textit{GitHub}. Modifications to files are collected in \textit{commits}, which can be seen as snapshots of code edits. Developers can commit small and large modifications, though often commit code before switching tasks or finishing a unit of work. \textit{We use commits, as elemental contributions in OSS, to quantify the activity of individual developers}.

Data on commits contributed to public projects on GitHub is made available and dynamically updated on the GH Archive database\footnote{\url{https://www.gharchive.org/}}. We use this database rather than querying data from the GitHub API. The next step is to assign GitHub accounts of authors to their commits. Commits themselves contain plaintext names and email addresses of authors, which do not correspond directly to GitHub accounts. For instance, a GitHub account user may contribute commits from multiple computers, each linked to git via a different name and email. Merging these identities under one developer is a crucial part in geolocating GitHub users, as the clearest information about their geographic location is provided on their GitHub profile page. We therefore applied a method from the software engineering research community to link email addresses and specific commits to GitHub user accounts (which we assume correspond to individual developers) \cite{montandon2019identifying} using the GitHub API: for each email address, we select a random commit made by that email address and query the GitHub API to retrieve the specific account login associated to the commit. In case the API cannot resolve the account using this first commit, we try three additional commits submitted using this email. As we are interested in active GitHub contributors, we consider all email addresses with at least 100 commits over the two years of 2019 and 2020. This corresponds to an average of nearly one commit per week. We note that our subsequent results replicate completely when applying a stricter threshold of at least 200 commits for inclusion. The results of this analysis are available upon request.

Having associated GitHub accounts to contributions, we access information about users via the GitHub API. In particular we access the \textit{location} and \textit{Twitter account} of individual users, when provided. Through the \textit{Twitter API}, one can also retrieve location information of Twitter users, when provided. A third way of gathering information on location is through email suffixes, belonging either to a country or institutions such as universities. We describe our method of geolocating developers in the following section.

%%%%%%%%

\subsection*{Geolocation}
Given a collection of active GitHub contributors and their account information, our goal is to infer the location for as many users as possible. We first focused on the raw location fields provided on GitHub user profiles. Reflecting on the common pitfalls of geocoding online profiles, for instance that users may give unreal or sarcastic locations (``the moon'') \cite{hecht2011tweets}, we selected the Bing Maps API. This API resolves multiple input languages (``Vienna'' and ``Wien'' refer to the same location) and can handle inputs at varying scales from country to geolocation precise to within meters.

In case a user did not share a location on GitHub, or the Bing API was not able to geolocate the string that the user did share, we check whether the user linked to a Twitter account. If so, we attempted to geolocate the Twitter account in a similar way, using the location field provided by the user on Twitter. In case we could not geolocate a user from their Twitter data, we considered the email suffixes under which they made commits. Email suffixes can suggest the location of individuals in two ways: by the country domain (i.e. .jp for Japan, .it for Italy) or by a university suffix. In the latter case we imitate previous work associating GitHub contributors to universities \cite{valiev2018ecosystem} using a list of universities, their locations and email domains maintained by Hipo\footnote{\url{https://github.com/Hipo/university-domains-list}}. We only infer user country from email suffix data.

Using this pipeline we could geolocate 587,852 active OSS contributors (out of 1,124,874 accounts with at least 100 commits) to at least the country level. 502,415 or 85\% user locations were identified from their GitHub account information alone. In other words, by considering email suffixes and Twitter data we could increase our pool of geolocated developers by 15\%. As country-identifying email domains and Twitter use vary significantly between countries, we share data on the number of users identified by each of the three methods in the iterative process. We note that a share of users could only be geocoded at the country level, for instance those classified using email-suffix data or giving only coarse geographic information in their location fields (i.e. ``Austria''). Specifically, we could infer subnational locations for 415,783 users. Code to replicate our data mapping pipeline is available at: \url{https://github.com/n1tecki/Geography-of-Open-Source-Software}.

\rowcolors{2}{gray!25}{white}
\begin{table}[!t]
\ra{1.0} 
\resizebox{\textwidth}{!}{%
\begin{tabular}{@{}rlrlllrl@{}}
\toprule
\multicolumn{1}{l}{} &
  \multicolumn{2}{c}{Sourceforge 2008} &
  \multicolumn{2}{c}{GitHub 2010} &
  \multicolumn{2}{c}{\textbf{GitHub 2021}} &
   \\ %\midrule
\multicolumn{1}{l|}{Rank} &
  Country &
  \multicolumn{1}{l|}{Share} &
  Country &
  \multicolumn{1}{l|}{Share} &
  Country &
  \multicolumn{1}{l|}{Share} &
  \multicolumn{1}{c|}{Rank Chg. vs. 2008} \\ \cmidrule(l){2-8} 
\multicolumn{1}{r|}{1} &
  United States &
  \multicolumn{1}{r|}{36.1} &
  United States &
  \multicolumn{1}{r|}{38.7} &
  United States &
  \multicolumn{1}{r|}{24.6} &
  \multicolumn{1}{c|}{-} \\ 
\multicolumn{1}{r|}{2} &
  Germany &
  \multicolumn{1}{r|}{8.1} &
  UK &
  \multicolumn{1}{r|}{7.7} &
  China &
  \multicolumn{1}{r|}{5.8} &
  \multicolumn{1}{c|}{↑ 4} \\ 
\multicolumn{1}{r|}{3} &
  UK &
  \multicolumn{1}{r|}{5.1} &
  Germany &
  \multicolumn{1}{r|}{6.2} &
  Germany &
  \multicolumn{1}{r|}{5.6} &
  \multicolumn{1}{c|}{↓ 1} \\ 
\multicolumn{1}{r|}{4} &
  Canada &
  \multicolumn{1}{r|}{4.2} &
  Canada &
  \multicolumn{1}{r|}{4.3} &
  India &
  \multicolumn{1}{r|}{5.4} &
  \multicolumn{1}{c|}{↑↑ 7} \\ 
\multicolumn{1}{r|}{5} &
  France &
  \multicolumn{1}{r|}{3.8} &
  Japan &
  \multicolumn{1}{r|}{3.9} &
  UK &
  \multicolumn{1}{r|}{5.0} &
  \multicolumn{1}{c|}{↓ 2} \\ 
\multicolumn{1}{r|}{6} &
  China &
  \multicolumn{1}{r|}{3.1} &
  Brazil &
  \multicolumn{1}{r|}{3.6} &
  Brazil &
  \multicolumn{1}{r|}{4.4} &
  \multicolumn{1}{c|}{↑↑ 6} \\ 
\multicolumn{1}{r|}{7} &
  Australia &
  \multicolumn{1}{r|}{2.7} &
  France &
  \multicolumn{1}{r|}{3.2} &
  Russia &
  \multicolumn{1}{r|}{4.3} &
  \multicolumn{1}{c|}{↑↑ 6} \\ 
\multicolumn{1}{r|}{8} &
  Italy &
  \multicolumn{1}{r|}{2.6} &
  Australia &
  \multicolumn{1}{r|}{3.1} &
  France &
  \multicolumn{1}{r|}{3.8} &
  \multicolumn{1}{c|}{↓ 3} \\ 
\multicolumn{1}{r|}{9} &
  Netherlands &
  \multicolumn{1}{r|}{2.5} &
  Russia &
  \multicolumn{1}{r|}{2.3} &
  Canada &
  \multicolumn{1}{r|}{3.8} &
  \multicolumn{1}{c|}{↓↓ 5} \\ 
\multicolumn{1}{r|}{10} &
  Sweden &
  \multicolumn{1}{r|}{2.0} &
  Sweden &
  \multicolumn{1}{r|}{2.2} &
  Japan &
  \multicolumn{1}{r|}{2.7} &
  \multicolumn{1}{c|}{↑↑ 5} \\ 
\multicolumn{1}{r|}{11} &
  India &
  \multicolumn{1}{r|}{1.9} &
   &
  \multicolumn{1}{l|}{} &
  South Korea &
  \multicolumn{1}{r|}{1.9} &
  \multicolumn{1}{c|}{↑↑↑ 14} \\ 
\multicolumn{1}{r|}{12} &
  Brazil &
  \multicolumn{1}{r|}{1.8} &
   &
  \multicolumn{1}{l|}{} &
  Netherlands &
  \multicolumn{1}{r|}{1.8} &
  \multicolumn{1}{c|}{↓ 3} \\ 
\multicolumn{1}{r|}{13} &
  Russia &
  \multicolumn{1}{r|}{1.6} &
   &
  \multicolumn{1}{l|}{} &
  Spain &
  \multicolumn{1}{r|}{1.8} &
  \multicolumn{1}{c|}{↑ 1} \\ 
\multicolumn{1}{r|}{14} &
  Spain &
  \multicolumn{1}{r|}{1.6} &
   &
  \multicolumn{1}{l|}{} &
  Poland &
  \multicolumn{1}{r|}{1.8} &
  \multicolumn{1}{c|}{↑ 2} \\ 
\multicolumn{1}{r|}{15} &
  Japan &
  \multicolumn{1}{r|}{1.3} &
   &
  \multicolumn{1}{l|}{} &
  Australia &
  \multicolumn{1}{r|}{1.8} &
  \multicolumn{1}{c|}{↓↓ 8} \\ 
\multicolumn{1}{r|}{16} &
  Poland &
  \multicolumn{1}{r|}{1.2} &
   &
  \multicolumn{1}{l|}{} &
  Sweden &
  \multicolumn{1}{r|}{1.2} &
  \multicolumn{1}{c|}{↓↓ 6} \\ 
\multicolumn{1}{r|}{17} &
  Belgium &
  \multicolumn{1}{r|}{1.2} &
   &
  \multicolumn{1}{l|}{} &
  Italy &
  \multicolumn{1}{r|}{1.2} &
  \multicolumn{1}{c|}{↓↓↓ 9} \\ 
\multicolumn{1}{r|}{18} &
  Switzerland &
  \multicolumn{1}{r|}{1.0} &
   &
  \multicolumn{1}{l|}{} &
  Ukraine &
  \multicolumn{1}{r|}{1.2} &
  \multicolumn{1}{c|}{New} \\ 
\multicolumn{1}{r|}{19} &
  Austria &
  \multicolumn{1}{r|}{0.8} &
   &
  \multicolumn{1}{l|}{} &
  Switzerland &
  \multicolumn{1}{r|}{1.2} &
  \multicolumn{1}{c|}{↓ 1} \\ 
\multicolumn{1}{r|}{20} &
  Denmark &
  \multicolumn{1}{r|}{0.8} &
   &
  \multicolumn{1}{l|}{} &
  Indonesia &
  \multicolumn{1}{r|}{1.0} &
  \multicolumn{1}{c|}{New} \\ 
\multicolumn{1}{r|}{21} &
  Singapore &
  \multicolumn{1}{r|}{0.8} &
   &
  \multicolumn{1}{l|}{} &
  Taiwan &
  \multicolumn{1}{r|}{0.8} &
  \multicolumn{1}{c|}{↑↑↑ 9} \\ 
\multicolumn{1}{r|}{22} &
  Finland &
  \multicolumn{1}{r|}{0.8} &
   &
  \multicolumn{1}{l|}{} &
  Colombia &
  \multicolumn{1}{r|}{0.8} &
  \multicolumn{1}{c|}{New} \\ 
\multicolumn{1}{r|}{23} &
  Norway &
  \multicolumn{1}{r|}{0.7} &
   &
  \multicolumn{1}{l|}{} &
  Argentina &
  \multicolumn{1}{r|}{0.7} &
  \multicolumn{1}{c|}{↑ 4} \\ 
\multicolumn{1}{r|}{24} &
  Mexico &
  \multicolumn{1}{r|}{0.7} &
   &
  \multicolumn{1}{l|}{} &
  Mexico &
  \multicolumn{1}{r|}{0.7} &
  \multicolumn{1}{c|}{-} \\ 
\multicolumn{1}{r|}{25} &
  South Korea &
  \multicolumn{1}{r|}{0.7} &
   &
  \multicolumn{1}{l|}{} &
  Norway &
  \multicolumn{1}{r|}{0.7} &
  \multicolumn{1}{c|}{↓ 2} \\ 
\multicolumn{1}{r|}{26} &
  Israel &
  \multicolumn{1}{r|}{0.6} &
   &
  \multicolumn{1}{l|}{} &
  Belgium &
  \multicolumn{1}{r|}{0.7} &
  \multicolumn{1}{c|}{↓↓↓ 9} \\ 
\multicolumn{1}{r|}{27} &
  Argentina &
  \multicolumn{1}{r|}{0.6} &
   &
  \multicolumn{1}{l|}{} &
  Denmark &
  \multicolumn{1}{r|}{0.7} &
  \multicolumn{1}{c|}{↓↓ 7} \\ 
\multicolumn{1}{r|}{28} &
  Hungary &
  \multicolumn{1}{r|}{0.6} &
   &
  \multicolumn{1}{l|}{} &
  Finland &
  \multicolumn{1}{r|}{0.6} &
  \multicolumn{1}{c|}{↓↓ 6} \\ 
\multicolumn{1}{r|}{29} &
  Romania &
  \multicolumn{1}{r|}{0.5} &
   &
  \multicolumn{1}{l|}{} &
  Vietnam &
  \multicolumn{1}{r|}{0.6} &
  \multicolumn{1}{c|}{New} \\ 
\multicolumn{1}{r|}{30} &
  Taiwan &
  \multicolumn{1}{r|}{0.5} &
   &
  \multicolumn{1}{l|}{} &
  Austria &
   \multicolumn{1}{r|}{0.6} &
  \multicolumn{1}{c|}{↓↓↓ 11} \\ \bottomrule
\end{tabular}%
}
\caption{Country shares of active OSS contributors on GitHub in 2021. We include the top 30 countries and compare our data with similar snapshots reported in previous work using Sourceforge (2008)~\cite{gonzalez2008geographic} and GitHub (2010, only top 10 available)~\cite{takhteyev2010investigating}. Across countries the distribution has become more uniform. South and East Asian and Latin American countries have seen the greatest relative increase in share of global OSS contributors.}
\label{country_summary}
\end{table}

\subsubsection*{Data Availability}
A primary goal of this work is to make geographic data on OSS developers accessible to researchers. We have uploaded both national and regional datasets to GitHub, available at: \url{https://github.com/johanneswachs/OSS_Geography_Data}. In order to protect user privacy, we only share geographically aggregated counts of active users. For example, one file includes the number of active developers located in each of the 50 US States in 2021. In particular we share data in CSV files comarping counts of active OSS contributors across Countries, European NUTS2 regions, and sub-national units (states/provinces) of the US, Japan, China, India, Russia, and Brazil.

\section*{Analysis and Results}

We begin by reporting summary statistics on the number of developers we located in various countries and regions. We highlight which countries host the most OSS developers both in raw terms and per capita. We compare our results with those from previous works published in 2008 and 2010, studying the geographic distribution of developers on the Sourceforge and GitHub platforms, respectively. We also present evidence that the count of active OSS developers correlates strongly with a variety of measures of development and quality of living indicators, above and beyond economic development. We replicate these findings at the regional scale using data from the European NUTS2 regions. We then present an analysis of the geographic concentration of OSS developers within countries, finding that they are in general highly concentrated.

\subsection*{International Comparison}
In Table \ref{country_summary} we report the top thirty countries, ranked by overall share of active OSS developers, and compare data from 2021 with snapshot data from previous work, carried out over 10 years ago. At first glance, our results indicate significant changes in the global distribution of OSS activity since 2010 \cite{gonzalez2008geographic,takhteyev2010investigating}. While North American and Western European countries are still leading locations for OSS, Asian, Eastern European, and Latin American countries are catching up. In Table \ref{country_pc} we report the top 50 countries by active OSS developers per 100k inhabitants. Here the top ranks are dominated by small and wealthy European countries.

\rowcolors{2}{gray!25}{white}
\begin{table}
\begin{tabular}{lllrrr}
\toprule
Rank &         Country & ISO2 &  Count Total Contributors &  Pop. (mm) &  Cont. / 100k \\
\midrule

1  &         Iceland &   IS &                       421 &         0.4 &                    105 \\
2  &     Switzerland &   CH &                      7197 &         8.6 &                     84 \\
3  &          Norway &   NO &                      4012 &         5.3 &                     76 \\
4  &          Sweden &   SE &                      7323 &        10.3 &                     71 \\
5  &         Finland &   FI &                      3813 &         5.5 &                     69 \\
6  &         Denmark &   DK &                      3906 &         5.8 &                     67 \\
7  &     Netherlands &   NL &                     10773 &        17.3 &                     62 \\
8  &          Canada &   CA &                     22269 &        37.6 &                     59 \\
9  &         Estonia &   EE &                       760 &         1.3 &                     58 \\
10 &      Luxembourg &   LU &                       324 &         0.6 &                     54 \\
11 &     New Zealand &   NZ &                      2642 &         4.9 &                     54 \\
12 &       Singapore &   SG &                      3102 &         5.7 &                     54 \\
13 &         Ireland &   IE &                      2531 &         4.9 &                     52 \\
14 &   United States &   US &                    144371 &       328.2 &                     44 \\
15 &  United Kingdom &   GB &                     29452 &        66.8 &                     44 \\
16 &       Australia &   AU &                     10337 &        25.4 &                     41 \\
17 &         Germany &   DE &                     33212 &        83.1 &                     40 \\
18 &         Austria &   AT &                      3276 &         8.9 &                     37 \\
19 &          France &   FR &                     22551 &        67.1 &                     34 \\
20 &         Belgium &   BE &                      3935 &        11.5 &                     34 \\
21 &          Israel &   IL &                      2488 &         9.1 &                     27 \\
22 &         Belarus &   BY &                      2532 &         9.5 &                     27 \\
23 &        Portugal &   PT &                      2802 &        10.3 &                     27 \\
24 &       Lithuania &   LT &                       748 &         2.8 &                     27 \\
25 &          Poland &   PL &                     10406 &        38.0 &                     27 \\
26 &         Czechia &   CZ &                      2805 &        10.7 &                     26 \\
27 &        Bulgaria &   BG &                      1755 &         7.0 &                     25 \\
28 &        Slovenia &   SI &                       492 &         2.1 &                     23 \\
29 &          Latvia &   LV &                       443 &         1.9 &                     23 \\
30 &           Spain &   ES &                     10593 &        47.1 &                     22 \\
31 &           Malta &   MT &                       112 &         0.5 &                     22 \\
32 &          Taiwan &   TW &                      4979 &        23.6 &                     21 \\
33 &     South Korea &   KR &                     10921 &        51.7 &                     21 \\
34 &         Hungary &   HU &                      1813 &         9.8 &                     18 \\
35 &         Croatia &   HR &                       742 &         4.1 &                     18 \\
36 &          Russia &   RU &                     25271 &       144.4 &                     18 \\
37 &       Hong Kong &   HK &                      1303 &         7.5 &                     17 \\
38 &         Ukraine &   UA &                      7204 &        44.4 &                     16 \\
39 &          Serbia &   RS &                      1039 &         6.9 &                     15 \\
40 &          Cyprus &   CY &                       168 &         1.2 &                     14 \\
41 &          Greece &   GR &                      1510 &        10.7 &                     14 \\
42 &        Slovakia &   SK &                       719 &         5.5 &                     13 \\
43 &           Japan &   JP &                     15706 &       126.3 &                     12 \\
44 &         Uruguay &   UY &                       435 &         3.5 &                     12 \\
45 &          Brazil &   BR &                     25891 &       211.0 &                     12 \\
46 &      Costa Rica &   CR &                       593 &         5.0 &                     12 \\
47 &           Italy &   IT &                      7204 &        60.3 &                     12 \\
48 &         Romania &   RO &                      1979 &        19.4 &                     10 \\
49 &         Namibia &   NA &                       260 &         2.5 &                     10 \\
50 &       Argentina &   AR &                      4332 &        44.9 &                     10 \\
\bottomrule
\end{tabular}
\caption{Countries ranked by number of OSS developers per capita, top 50.}
\label{country_pc}
\end{table}

\begin{figure}[!t]
\centering
\includegraphics[width=\textwidth]{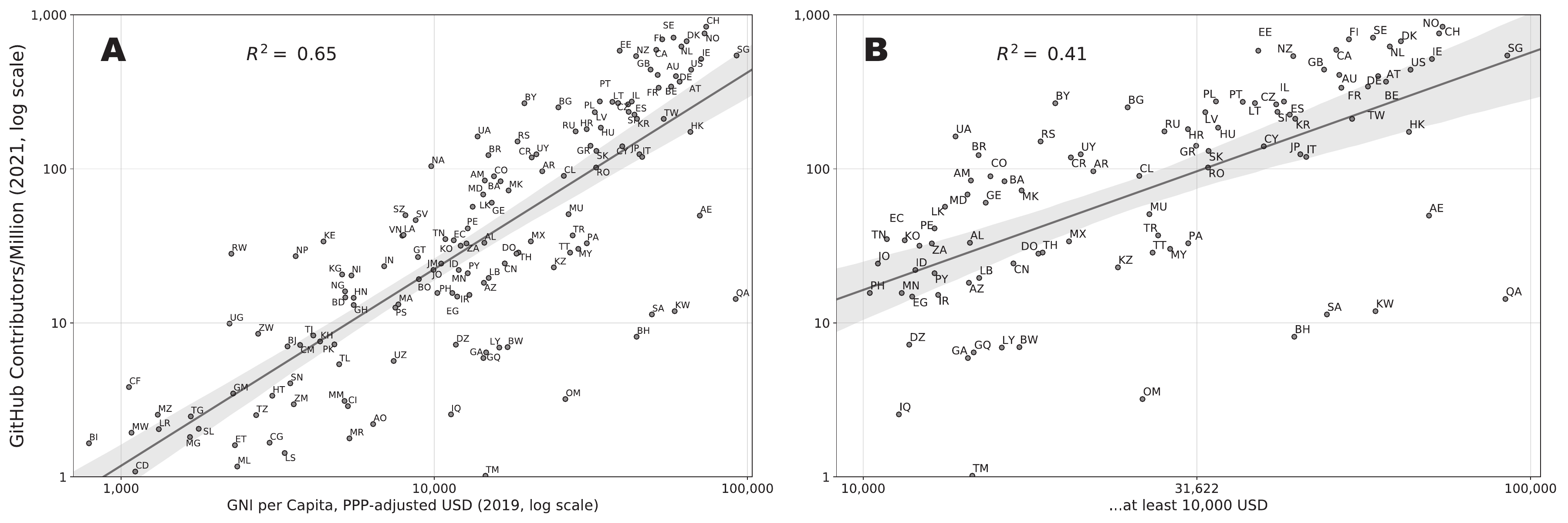}
\captionsetup{width=\linewidth}
\caption{Country economic development (PPP-adjusted GNI per capita) and OSS contributors per capita, log-log scale. A) We observe a strong relationship between economic development and OSS activity, though some countries deviate significantly from the trend. B) Zooming in on wealthier countries only, the relationship weakens significantly, suggesting other factors play an important role in OSS activity. For sake of visualization, we exclude countries with fewer than 1 million inhabitants or less than 1 contributor per million inhabitants.}
\label{fig:econ_vs_oss}
\end{figure}

A more interesting ranking of national activity in OSS would take into account both the population of each country and its level of economic development. In Figure \ref{fig:econ_vs_oss} we present the relationship between income per capita, sourced from the World Bank, and the number of active OSS developers per million inhabitants, both on logarithmic scales. We exclude countries with a population of less than one million people for the sake of visualization. The regression fit explains roughly two-thirds of the variance among all countries, but only 40\% for countries with an income per capita of at least \$10,000. Countries above the regression line have more OSS developers per capita than expected for their level of economic development, while those below have less. Ukraine, Belarus, Namibia, Brazil, Bulgaria, and Estonia have more OSS activity than expected, while oil-rich states like Qatar, Kuwait, and Saudi Arabia are OSS laggards.

What explains these residuals? The ability and decision to contribute to OSS projects is likely a complex and multifaceted process \cite{gerosa2021shifting}, but we can compare the relative importance of various structural factors in a regression framework. Beyond the broad economic development of a country, measured by income per capita, internet penetration \cite{ITU2019} likely plays an important role. The UN's Human Development Index (HDI) offers a broad measure of social development, including access to education and health services - likely important upstream factors facilitating OSS contributions \cite{hdi2019human}. 

\rowcolors{2}{white}{white}
\begin{table}[]
\centering
\ra{1.1}  
\begin{tabular}{@{}lccc@{}}
\toprule
Country Feature & Spearman $\rho$  & p-Value & Observations         \\ \midrule
PPP-adjusted GNI per capita (2019)                   & 0.79 & $<0.01$         & 176        \\ 
Human Development Index (HDI, 2019)                   & 0.86 & $<0.01$         & 178       \\ 
WVS share ``most people can be trusted''                    & 0.68 & $<0.01$         & 75        \\ 
Index of Public Integrity (IPI, 2019)                & 0.87 & $<0.01$         & 117        \\ 
Economic Complexity Index (ECI, 2019)                   & 0.82 & $<0.01$         & 175        \\ 
Deep-learning/AI publications/capita                   & 0.67 & $<0.01$         & 183        \\ 
Internet Penetration (2019)                   & 0.78 & $<0.01$         & 180        \\ 
\bottomrule
\end{tabular}%
\caption{Spearman rank correlations between country-level social and economic development indicators and active OSS contributors per capita in 2021.}
\label{tab:otherPlatforms}
\end{table}

One distinguishing aspect of OSS, compared to many other kinds of knowledge-intense activities, is that OSS outputs are essentially public goods - goods that can be used by anyone. Though a thorough review of the economics of OSS \cite{lerner2002some,sahay2019free} and individual motivations for contributing \cite{gerosa2021shifting} is beyond the scope of this article, we do expect that OSS activity will be higher where people are more inclined to contribute to public goods. For instance, people living in areas with high levels of generalized trust, that is to say where individuals are more likely to report that in general ``most people can be trusted'', are known to be more likely to contribute to public goods \cite{rothstein2005all}. We use data from the most recent wave of the World Values Survey to measure this concept of generalized trust \cite{haerpferm}. Another feature of countries that strongly correlates with individual propensity to contribute to public goods is quality of government. We therefore also relate OSS outcomes to the Index of Public Integrity (IPI), a measure of the quality of public institutions in a country \cite{mungiu2016measuring}.

\begin{table}[!b] \centering
\begin{tabular}{@{\extracolsep{5pt}}lccccc}
\\[-1.8ex]\hline
& \multicolumn{5}{c}{Active GitHub Contributors per Million Inhab. (log, 2021)} \
\cr
\\[-1.8ex] & (1) & (2) & (3) & (4) & (5) \\
\hline \\[-1.8ex]
PPP GNI per Cap. ('000 USD, 2019) & 0.017$^{*}$ & -0.007$^{}$ & -0.002$^{}$ & 0.004$^{}$ & -0.010$^{*}$ \\
  & (0.009) & (0.007) & (0.009) & (0.008) & (0.006) \\
Internet Penetration (\% of Pop., 2019)& 0.043$^{***}$ & 0.002$^{}$ & 0.029$^{***}$ & 0.026$^{***}$ & -0.003$^{}$ \\
  & (0.005) & (0.008) & (0.005) & (0.006) & (0.009) \\
Population (log, 2019) & -0.145$^{***}$ & -0.071$^{}$ & -0.016$^{}$ & -0.143$^{**}$ & -0.038$^{}$ \\
  & (0.052) & (0.044) & (0.046) & (0.056) & (0.057) \\
Human Development Index (2019) & & 11.709$^{***}$ & & & 9.327$^{***}$ \\
  & & (1.528) & & & (1.553) \\
Index of Public Integrity (2019) & & & 0.704$^{***}$ & & \\
  & & & (0.127) & & \\
Economic Complexity Index (2019) & & & & 0.962$^{***}$ & 0.683$^{***}$ \\
  & & & & (0.184) & (0.153) \\
\hline \\[-1.8ex]
 Observations & 174 & 173 & 115 & 150 & 149 \\
 Adjusted $R^2$ & 0.631 & 0.747 & 0.819 & 0.740 & 0.804 \\
 Residual Std. Error & 1.191 & 0.986 & 0.788 & 1.016 & 0.882  \\
 F Statistic & 81.3$^{***}$  & 175.2$^{***}$  & 195.8$^{***}$  & 142.4$^{***}$  & 155.6$^{***}$  \\
\hline
  & \multicolumn{5}{r}{$^{*}$p$<$0.1; $^{**}$p$<$0.05; $^{***}$p$<$0.01} \\
\end{tabular}
\caption{Regression models (1-5) relating country-level counts of GitHub contributors per million inhabitants (log-transformed) and socio-economic indicators. While income and internet penetration alone account for nearly two-thirds of variance in OSS activity (1), human development (2), quality of political institutions (3), and economic complexity (4) significantly improve model fit above and beyond that baseline. A combined model (5) explains over 80\% of variance. We report robust standard errors.}
\label{tab:country_regressions}
\end{table}

Lastly, we consider the overall economic focus and specialization of a country. As noted before, oil-producing states seem to have fewer OSS contributors relative to their economic development. The extent to which countries specialize in more complex, coordination-intensive industries likely correlates significantly with OSS activity. To capture this aspect, we consider the Economic Complexity Index (ECI) \cite{hidalgo2021economic}, a measure of the sophistication of a country's export profile. To measure national sophistication in a frontier and software-adjacent field, we use a count the number of academic research preprints on AI/deep-learning published by each country in the last decade \cite{klinger2021deep}. 

Each of these social, political and economic features has a strong correlation with the local intensity of OSS contributors. We report the Spearman rank correlation coefficient ($\rho$) of each variable with OSS contributors per capita in Table \ref{tab:otherPlatforms}. We report a full correlation matrix and basic summary statistics in the appendix. Though it is not possible to disentangle the cause and effect relationships between these variables using observational data, we can observe how some of these variables mediate each other, and how they can explain a greater share of inter-country variance in OSS contributors. To do so, we regress the (log-transformed) number of OSS contributors per million inhabitants on a selection of these variables, reporting the results of several alternative specifications fit using ordinary-least-squares (OLS) in Table~\ref{tab:country_regressions}, reporting robust standard errors and measuring the goodness of model fit using adjusted R-squared.

Our baseline model, including per capita income, internet penetration, and population, explains around two-thirds of variance as measured by adjusted R-squared. Adding HDI, IPI, or ECI as individual features increases the variance explained by over 10\%. All three variables have a significant positive relationship with great OSS activity in a country. A model combining the baseline model with HDI and ECI accounts for over 80\% of variance in OSS activity. In general we observe both a statistical significant relationship between these socio-economic factors in our models, and a large increase in the quality of the model fit when including them. Given that our features have significant pairwise correlations, we test for issues of multicollinearity which may influence our estimates using Variance Inflation Factor (VIF) tests for each specification. The results, reported in the appendix, indicate moderate levels of correlation within standard bounds.

These models suggest that OSS activity is not merely the by-product of an advanced economy, but also depends to a significant degree on social, educational, and political institutions and the degree of technological sophistication in a country. These results shed light on the residuals of the simple bivariate relationship between economic development and OSS activity reported in Figure\ref{fig:econ_vs_oss}. To take an extreme example, though Estonia and Bahrain have comparable average income, the density of OSS contributors in Estonia is nearly two orders of magnitude greater than that of Bahrain. Much of this difference is captured by variation in, for example human development or economic complexity. But even Latvia and Lithuania, geographically and historically linked to Estonia, have significantly fewer OSS developers - suggesting rich nuance in the variation of OSS activity and scope for specific policy interventions, to be discussed later in the paper. First, however, we will zoom in on the sub-national level. As we will see in the following section, geographic variance in OSS activity becomes more difficult to explain with macro indicators at finer spatial scales.

\section*{Regional Variation}
Comparing OSS activity between countries indicates that it has spread internationally to a significant extent in the last ten years. However, we know little about the distribution of activity within countries. As mentioned above, while most prior work has focused on international comparisons, Takhteyev and Hilts \cite{takhteyev2010investigating} reported data on local clusters in their work from 2010. They estimated that 7.4\% of \textit{global} contributors were in the San Francisco Bay Area, suggesting an immense local concentration of OSS activity. In this section we explore the local distribution of OSS developers in various countries. We focus first on European NUTS2 regions. We again relate socio-economic features to OSS activity, replicating our international findings at the regional scale. We also report data on the top US metropolitan statistical areas. We then consider the \textit{concentration} of developers within multiple countries including the EU, US, China, India, Japan, and Brazil. We find that OSS activity is significantly concentrated relative to the distribution of the general population in all countries we examine, though with significant heterogeneity. Repeating the calculation using university educated workers or workers employed in high-tech. fields instead of OSS contributors, we find far lower levels of concentration.

\subsection*{European Regions}

\begin{figure}[!h]
\centering
\includegraphics[width=0.9\textwidth]{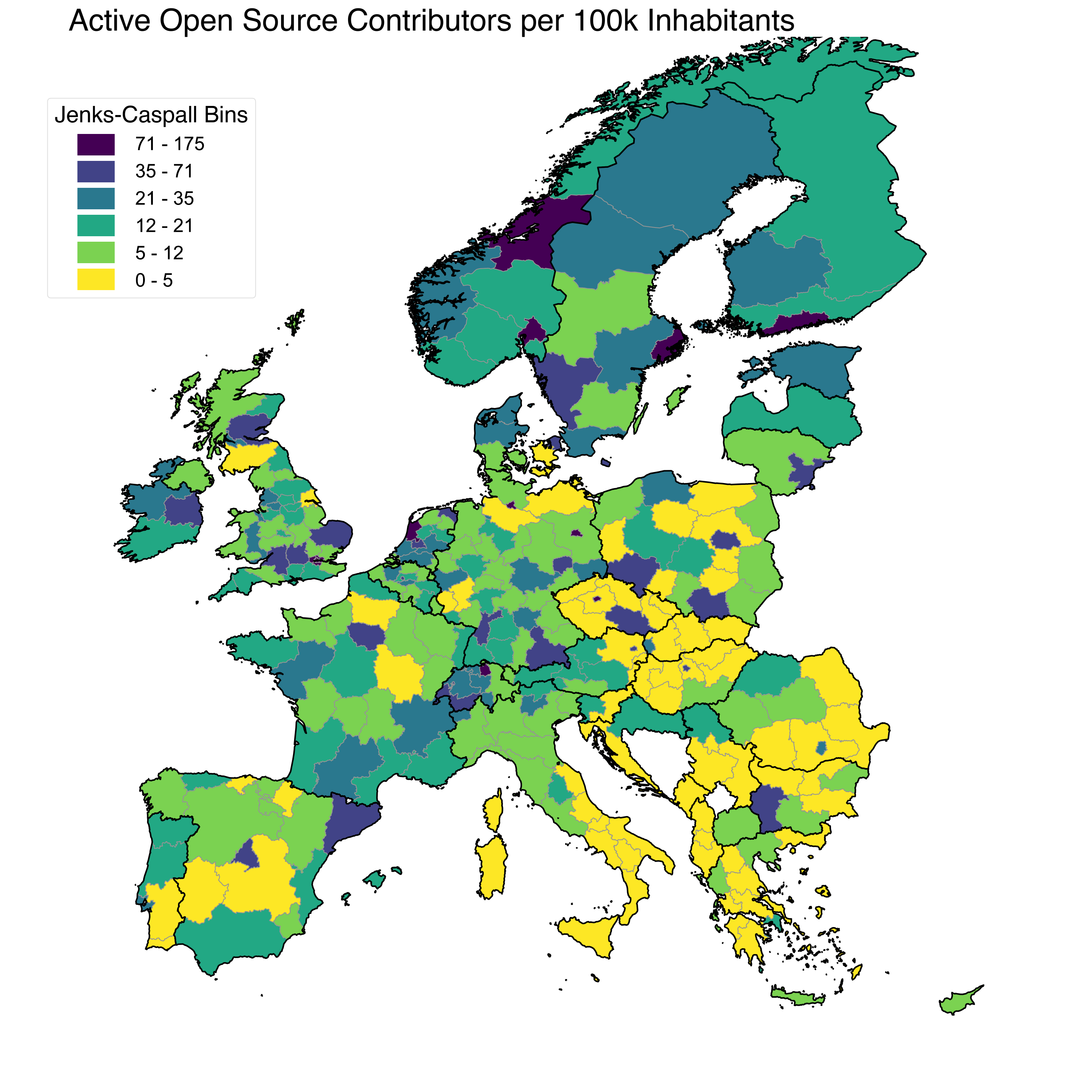}
\captionsetup{width=\linewidth}
\caption{Active OSS developers concentrations in early 2021 per 100,000 inhabitants, NUTS2 regions. We observe significant within country variation.}
\label{fig:nutsmap}
\end{figure}

We zoom in on European NUTS2 regions. These regions are especially useful because we can compare regions from multiple countries with generally consistent statistics, sourced from Eurostat. In particular we use the regions defined in 2016. In Figure \ref{fig:nutsmap} we map the number of OSS developers per 100,000 inhabitants by European NUTS2 region using Jenks-Caspall bins. We note that the five London NUTS2 regions are merged into a single unit because developers tended to refer to their location as London rather than ``Inner London''. We can observe several patterns. First we note that major hubs such as London, Amsterdam, Berlin, Prague, Zurich, Hamburg, Helsinki, Oslo and Stockholm tend to have significant OSS presence. Second, we can see significant variation between regions within countries. Some countries have regions in  both the lowest and highest valued bins. In other countries, such as Italy, Spain, and France, the distribution seems to be more uniform. We will return to an investigation of the within-country spatial concentration of developers later.

Knowing the locations of OSS developers at the regional levels, we can attempt to replicate our earlier findings about the relationship between socio-economic development indicators and OSS activity. Again the goal is to show that OSS activity is related to more than just economic development outcomes, albeit this time at a sub-national scale. On top of a baseline set of features including internet penetration, GDP per capita, population and density, we consider the relationships between various indicators of social and technological development and OSS activity. In particular we consider general levels of social trust, measured by the European Values Survey \cite{EVS2017}, R\&D Spending per capita, share of workers employed in high-tech industries, number of patents per 100k inhabitants (sourced from the OECD REGPAT database \cite{maraut2008oecd}), and share of the prime working-age population with tertiary education. Unless otherwise noted data are sourced from the 2017 QoG Basic dataset \cite{dahlberg2021quality}. When data on any feature is only available at the coarser NUTS0 (country) level, we impute from the country level to the regions. Finally we also consider the number of patents filed in the Electrical Engineering (EE) sector in case the relationship between OSS activity and closed-source innovation is heterogeneous across sectors. The WIPO IPC Technology Concordance Tables categorize patents into five fields at its coarsest level, of which EE is most directly related to software\footnote{The other categories are: Instruments, Chemistry, Mechanical Engineering, and Other.}.

We would like to model the relationship between these indicators of socio-economic development and OSS activity per capita at the NUTS2 level, observing both statistical significance and the share of variance explained. We again constructed a baseline model (including internet penetration, GDP per capita, population, and population density) and added the other features one at a time. However, when modeling outcomes of geographic units at finer spatial scales, it is important to consider spatial correlations in variables of interest \cite{cliff1981spatial}. Such correlations violate OLS assumptions about the independence of the error term and can introduce bias. To test for spatial autocorrelation in our data and modeling set up, we first calculated Moran's I on the distribution of OSS developers per 100k inhabitants (logged) across the NUTS2 regions. Two regions are considered adjacent if they share any border, including corners. We found evidence of significant spatial correlation ($I=0.42$, $p<.001$).

This suggests that linear regressions fit by OLS may suffer from bias introduced by spatial autocorrelations. To test this, we estimated these models and applied diagnostic tests (Moran's I of the residuals and Lagrange Multiplier tests \cite{anselin1988lagrange}). These tests, reported in the Appendix, confirmed that our estimates are significantly biased by spatial autocorrelation (i.e. the p-value of Moran's $I$ calculated on the residuals was always below 0.1, and below .05 for a majority of the models). We therefore used a general method of moments (GMM) modeling approach developed by Arraiz et al. that corrects this bias and adjusts for spatial correlation in both the dependent variable and independent variables \cite{arraiz2010spatial}. We used the Pysal implementation of this model in the Python programming language \cite{rey2010pysal}.

\begin{table}[] \centering
\begin{tabular}{@{\extracolsep{5pt}}lccccccc}
\\[-1.8ex]\hline
& \multicolumn{7}{c}{Active GitHub Contributors/100k Inhab. European NUTS2 (log, 2021)} \
\cr
\\[-1.8ex] & (1) & (2) & (3) & (4) & (5) & (6) & (7) \\
\hline \\[-1.8ex]
  Internet Penetration & 0.010$^{**}$ & 0.005$^{}$ & 0.012$^{***}$ & 0.012$^{***}$ & 0.012$^{**}$ & 0.011$^{**}$ & 0.004$^{}$ \\
  ---(\% of Pop. 2017)  & (0.004) & (0.005) & (0.004) & (0.004) & (0.004) & (0.004) & (0.004) \\
 GDP per Cap. & 0.600$^{***}$ & 0.609$^{***}$ & 0.012$^{}$ & 0.367$^{**}$ & 0.248$^{}$ & 0.269$^{}$ & 0.352$^{**}$ \\
 ---(log Eur, 2017)   & (0.140) & (0.154) & (0.256) & (0.147) & (0.176) & (0.202) & (0.151) \\
 Population  & 0.143$^{*}$ & 0.178$^{}$ & 0.356$^{***}$ & 0.166$^{*}$ & 0.063 & 0.089$^{}$ & 0.207$^{**}$ \\
 ---(log, 2020)   & (0.084) & (0.113) & (0.131) & (0.092) & (0.086) & (0.085) & (0.081) \\
 Population Dens. & 0.043$^{*}$ & 0.036$^{}$ & 0.044$^{}$ & -0.018$^{}$ & 0.055$^{**}$ & 0.050$^{*}$ & 0.030$^{}$ \\
 ---(log, 2017) & (0.023) & (0.024) & (0.029) & (0.022) & (0.027) & (0.028) & (0.022) \\
  EVS Trust  & & 0.315$^{**}$ & & & & & \\
 ---(2017) & & (0.151) & & & & & \\
 R\&D Spend. per Cap. & & & 0.113$^{**}$ & & & & \\
 ---(log, 2017)  & & & (0.051) & & & & \\
 \% Empl. High-Tech & & & & 0.083$^{***}$ & & & \\
 ---(2019/20) & & & & (0.011) & & & \\
 Patents Elec-Eng./100k & & & & & 0.058$^{***}$ & & \\
 ---(log, 2017)  & & & & & (0.022) & & \\
 Patents/100k  & & & & & & 0.047$^{*}$ & \\
 ---(log, 2017)  & & & & & & (0.026) & \\
 \% with Tertiary Edu. & & & & & & & 0.021$^{***}$ \\
 ---(2019/20) & & & & & & & (0.002) \\
Lambda & 0.099$^{***}$ & 0.114$^{***}$&0.073$^{***}$& 0.083$^{***}$ & 0.078$^{***}$ &0.084$^{***}$ &0.071$^{***}$ \\
---(est. spatial dep.) & (0.014) &(0.012) & (0.015) & (0.015) & (0.013) & (0.014) & (0.014) \\

\hline \\[-1.8ex]
 Observations & 276 & 198 & 258 & 262 & 258 & 258 & 276 \\
 Pseudo-$R^2$ & 0.392 & 0.429 & 0.417 & 0.509 & 0.411 & 0.399 & 0.536 \\
\hline
& \multicolumn{7}{r}{$^{*}$p$<$0.1; $^{**}$p$<$0.05; $^{***}$p$<$0.01} \\
\end{tabular}
\caption{GMM spatial regression models (1-7) (\cite{arraiz2010spatial}) relating EU NUTS2 counts of GitHub contributors per 100k inhabitants (log-transformed) and socio-economic indicators. Income, population, and internet penetration account for just over one third of variance in OSS activity (1). Social trust (2), R\&D spending (3), employment in high tech sectors (4), innovation activity (5,6), and higher education (7) all explain additional variance above this baseline (from 1 to 14\%). In each model the spatial autoregressive term lambda is positive and significant, indicating a positive adjacency relationship: neighboring regions tend to have similar levels of OSS activity even accounting for the features in each model.}
\label{tab:nuts_regressions}
\end{table}

We report the results of our GMM regression analysis in Table \ref{tab:nuts_regressions} with the (log-transformed) number of active GitHub contributors in a region per 100k inhabitants as the dependent variable. The baseline model again indicates that economic development and internet access have a significant positive relationship with OSS activity in regions. However, at this spatial scale the model has a significantly less accurate fit (pseudo $R^{2} \approx .39$) than a similar model predicting OSS activity at the country level. This observation applies also to the feature-rich models. Though generalized trust, R\&D spending per capita, share of employment in high-tech industries, share of population with tertiary education, and patents are significant predictors of greater OSS activity, the overall model fit only improves somewhat when adding these features. Only when we include share of the working age population with tertiary education or share of workers employed in high-tech sectors does the variance explained exceed 50\%. While factors like the presence of technologically advanced industries and an educated workforce clearly relate to OSS activity, it seems that at the regional level, more idiosyncratic forces determine local participation in OSS.

The relationships between the two patenting variables and OSS activity also merit comment. Patents are awarded to protect intellectual property and to block the uncompensated use of a creator's ideas. In a naive sense, patenting would appear to be a substitute for activity in open source and the creation of public goods. In practice, however, we see that patenting in electrical engineering, and to lesser extent patenting in all fields, has a significant positive relationship with OSS activity in regions. If OSS activity were to crowd-out patenting, we would expect to see the opposite relationship. This suggests that OSS plays a complementary role in the innovation process, likely via knowledge spillovers discussed earlier. Hybrid outcomes are also possible, in which software that accompanies a proprietary product is made open source. More work is needed to understand the potential impact of policy interventions on the relationship between open source activity and patenting. This finding also suggests how OSS activity can serve as a proxy of useful skills cognitively close to those involved in closed-source innovation \cite{boschma2005proximity}.

We have seen that in regions, OSS activity is positively related with economic development, activity in technology intensive sectors, and the presence of an educated and trusting population. At the same time, models including these factors fail to explain over half of the observed variance in OSS activity between regions in Europe. This suggests that while a place can have the right ingredients for an OSS hotspot, for instance a wealthy, well-educated, and digitally connected populace and still fall short. These findings motivate our analysis of the concentration of OSS activity within countries, which will demonstrate that this variance is also large in size.

\subsection*{US Metropolitan Statistical Areas}
Before turning to an analysis of within country concentration, we briefly present data on the distribution of developers in US metropolitan areas. Our reasons for this detour are twofold: first we can estimate the share of the global developer population in the San Francisco Bay Area and compare it with previous estimates from 2010. Second, the results indicate that concentration is also high in the US. While US states often have populations similar to European countries, and US cities often contain multiple counties, the US Census bureau defines Metropolitan Statistical Areas (MSAs) to highlight urban agglomerations. In Table~\ref{tab:usmsa} we report the top 10 MSAs with at least one million inhabitants by the number of developers per capita. The two leading regions are San Francisco and Silicon Valley, while Seattle has a similar density. A back of the envelope calculation suggests that together the two Bay Area regions account for about 3.7\% of all OSS developers world wide (summing 4,587 and 10,702, and dividing by 415,783, the number of OSS developers we could locate at a subnational level). This is around half of an estimate of the share of OSS developers in Silicon Valley from 2010 of 7.8\% \cite{takhteyev2010investigating}. We report a table of the top 50 MSAs by contributors per capita including those with smaller populations in the Appendix, noting that many university towns appear in prominent positions.

\begin{table}
\begin{tabular}{lrrr}
\toprule
                                       MSA Name &  Count Contributors &  Population &  Contributors/100k \\
\midrule
              San Jose-Sunnyvale-Santa Clara, CA &                4,587 &     1,990,660 &                230 \\
              San Francisco-Oakland-Hayward, CA &               10,702 &     4,731,803 &                226 \\
                    Seattle-Tacoma-Bellevue, WA &                8,830 &     3,979,845 &                221 \\
                          Austin-Round Rock, TX &                3370 &     2,227,083 &                151 \\
            Portland-Vancouver-Hillsboro, OR-WA &                2,751 &     2,492,412 &                110 \\
                 Boston-Cambridge-Newton, MA-NH &                5,221 &     4,873,019 &                107 \\
                     Denver-Aurora-Lakewood, CO &                2,555 &     2,967,239 &                 86 \\
                                    Raleigh, NC &                1,159 &     1,390,785 &                 83 \\
                             Salt Lake City, UT &                 989 &     1,232,696 &                 80 \\
          New York-Newark-Jersey City, NY-NJ-PA &               11,579 &    19,216,182 &                 60 \\
                 \bottomrule
          \end{tabular}

\caption{Top 10 US Metropolitan Statistical Areas with at least one million inhabitants ranked  by active GitHub developers per capita.}
\label{tab:usmsa}
\end{table}

\subsection*{Concentration}
As we have noted above, many kinds of knowledge-intensive activities are generally known to cluster in specific areas within countries. One of the primary goals of our study is to examine the extent to which this is true of OSS. Though it may be especially conducive to remote collaboration and decentralization, Figure \ref{fig:nutsmap} provides qualitative evidence that these aspects of OSS development do not outweigh the tendency of knowledge-intensive activities to cluster. We now introduce a measure to quantify this phenomenon, and compare the degree of geographic concentration of OSS activity in countries. 

There are many measures of the dispersion or concentration of people or things across geographic regions \cite{ellison1997geographic}. As we are interested in comparing the relative concentration of OSS developers between countries, we need a measure which considers population heterogeneity between regions within countries. For example, in a hypothetical country with two regions, A and B containing 80\% and 20\% of the country's population, respectively, an 80\%-20\% distribution of OSS developers between regions A and B should not be interpreted as concentration.

Measures like the Herfindahl Index depend on the number of regions, while the Ellison-Glaeser measure is sensitive to variance in population between regions \cite{ellison1997geographic}, and Gini-like measures quantify inequality, which is distinct from concentration. The \textit{Adjusted Geographic Concentration} ($AGC$), developed by the OECD \cite{spiezia2003measuring}, measures concentration that is comparable between countries with different numbers of regions and different distributions of the underlying population between them \cite{rovolis2006ethnic}. 

Consider a country $C$ with a population $P$ split into $N$ regions. The regions $i \in \{1, 2, \ldots N \}$ have shares of the population $p_{i}$. Denoting by $m_{i}$ the share of OSS developers in a country living in the region $C_{i}$, we define the \textit{Geographic Concentration} (GC) of developers in country $C$ as:

$$ GC(C) = \sum _{i \in C}^{N} \left| m_{i} - p_{i} \right|.  $$

This measure sums the absolute differences in shares between the general population and the subpopulation of interest (in our case, active OSS contributors). This statistic tends to underestimate concentration in regions with a larger share of the population, and the validity of comparisons between countries with different numbers of regions is unclear. To address these issues the $GC$ is usually scaled by its maximum possible value in each country: specifically, by the value it would take if all OSS developers were located in the least populated region of the country. In our previous notation:

$$GC_{max}(C) = (1 - p_{min}) + \sum _{i \in C, p_{i} \neq min}^{N} p_{i} =  2 (1-p_{min}) $$
Dividing $GC(C)$ by $GC_{max}(C)$, we obtain the \textit{Adjusted Geographic Concentration} (AGC) of a country $C$:

$$AGC(C) = \dfrac{GC(C)}{GC_{max}(C)} $$

The AGC varies between 0 and 1: a country in which the population of OSS developers is distributed in precisely the same proportions across region as the population, would have an AGC score of 0. A country in which all of OSS developers live in the region with the smallest population would have an AGC score of 1.

\begin{figure}[!t]
\centering
\includegraphics[width=\textwidth]{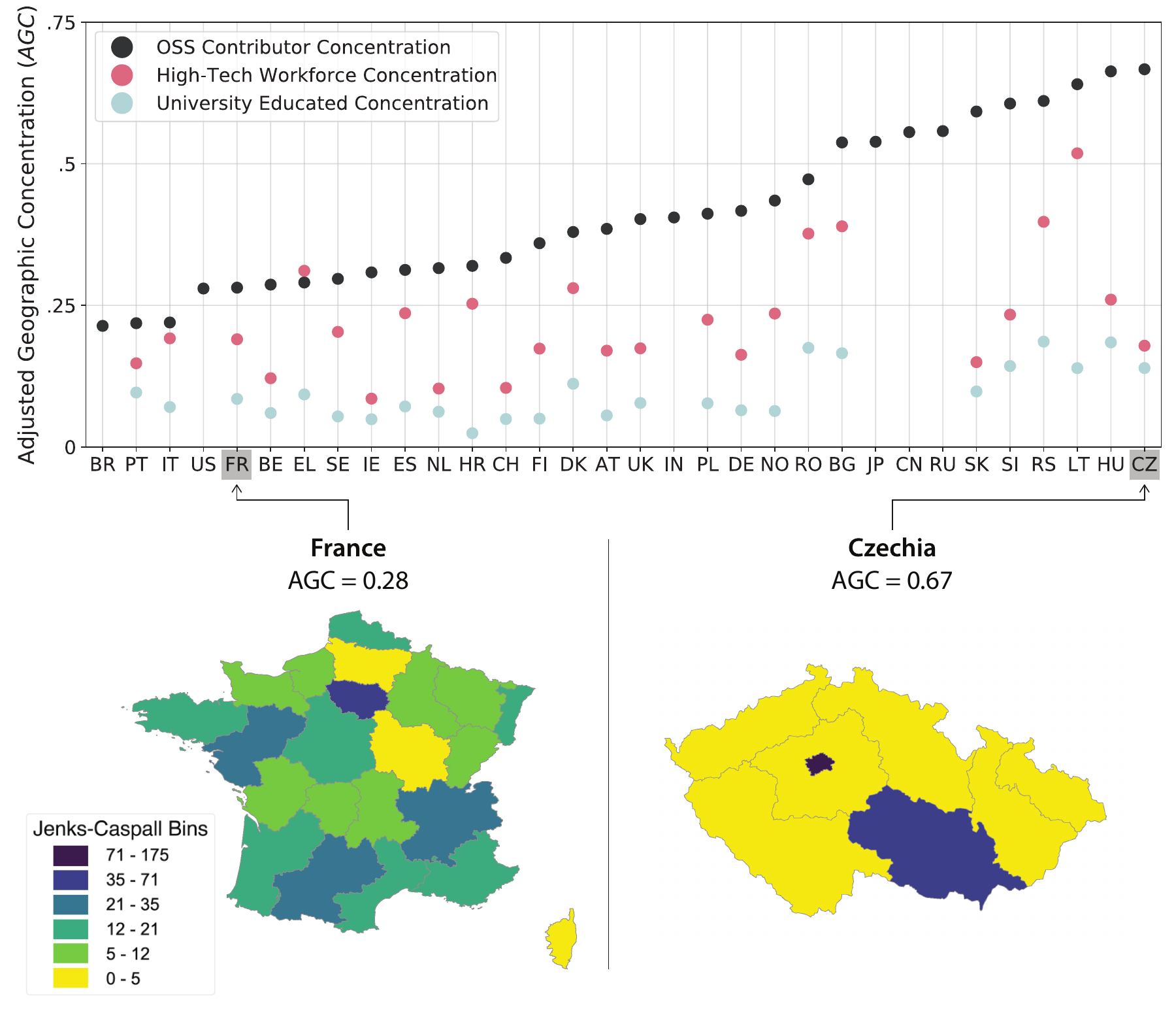}
\captionsetup{width=\linewidth}
\caption{OSS contributor regional concentration within countries, measured using Adjusted Geographic Concentration ($AGC$). A score of zero indicates that OSS contributors are distributed across a country's regions in proportion to population. $AGC$ hypothetically equals 1 if all OSS contributors are concentrated in the least populated region of the country. For European countries with NUTS regions we compare the $AGC$ of OSS contributors with the $AGC$ calculated for university educated workers and people employed in high-tech sectors. Below: Examples of countries with relatively low (France) and high (Czechia) OSS $AGC$ scores. Darker NUTS2 regions indicate more OSS contributors per capita, with bins as in Figure \ref{fig:nutsmap}.}
\label{fig:conc}
\end{figure}

We calculated the AGC score for various countries, including European countries with at least 2 NUTS2 regions, and the US (states + DC), China (provinces and municipalities, excluding Hong Kong and Taiwan), India (states and union territories), Russia (federal subjects) and Brazil (federal states + the Federal District). We report our estimates of developer concentration by country in Figure \ref{fig:conc}. 

We see that in all countries we examine, OSS development is concentrated regionally, relative to the general population. There is however significant variation between countries. For instance we can say that Brazilian, Portuguese and Italian OSS contributors are more evenly distributed amongst regions in those countries, than developers in Czechia, Hungary and Lithuania.

This analysis, however, does not make it clear how concentrated OSS developers are compared to other kinds of knowledge workers. In general regional statistics on such workers are not internationally comparable. However, among the European NUTS countries, we can make this comparison. We therefore recalculated the $AGC$ of each country in this group, substituting the share of workers in high-tech sectors and with tertiary education, respectively, for OSS contributors in the calculation. If OSS contributors are more dispersed in a country than the university educated or high-tech workforce, we would expect the $AGC$ to be higher under these alternative specifications. 

In Figure~\ref{fig:conc} we observe the opposite effect: OSS contributors are significantly \textit{more} concentrated in particular regions than either university educated or high-tech workers. This finding holds with remarkable regularity across the countries we analyze (only Greece is an exception). We provide the full table in the appendix, see Table \ref{tab:nuts_agc}. These estimates of concentration also provide useful perspective for the previous more global analysis. The $AGC$ scores for workers with higher education vary between .02 and .19 (mean: .09, stdev.: .05), for workers in high-tech sectors between .1 and .52 (mean: .23, stdev.: .10), and for OSS contributors in European countries between .22 and .67 (mean: .41, stdev.: . 14). 

The $AGC$ of OSS developers in all countries in our sample exceeds the average $AGC$ of high-tech workers in European countries. Overall, these results present strong evidence that OSS developers cluster to a significant degree in all countries in our analysis. While public policy promoting OSS tends to be made at the national level \cite{blind2021impact}, this analysis suggests that there is ample room for the regional promotion of OSS. We revisit this point in the following section.

\section*{Conclusions and Implications}

In this paper we presented an analysis of a novel dataset on the geography of open source software developers. We found that while the overall share of active developers has become more evenly distributed between countries, within-country regional differences remain strong. These heterogeneities are likely to persist: the social and economic spillovers of local OSS activity seem to be self-reinforcing. If indeed OSS is a driver of, and not merely a proxy for innovation outcomes, we need to better understand the role of actors such as universities, research institutes, and governments in promoting OSS activity \cite{secundo2017intellectual,nagle2019government}, for example in a mission-oriented context \cite{wanzenbock2020framework}. While most policy supporting OSS activity focuses on the national level, our findings suggest that local and regional policy may be more appropriate.

Indeed, although digitalization facilitates collaborations across distances \cite{forman2019digital}, continued regional clustering suggests that location matters as much as ever. The network effects present in places such as Silicon Valley are strong enough to overcome other obstacles including higher tax rates and cost of living. In software knowledge spillovers and transfers still happen locally and within firms \cite{wu2018all}. The developers in our dataset were geocoded in early 2021, roughly one year after the Covid-19 crisis went global. It remains to be seen whether proliferation of remote work and decentralization will be reflected in a change in the geographic distribution of OSS developers. If knowledge workers are going to permanently decamp to smaller cities in the post-Covid era, one would except to see the first signs of this in OSS with its advanced infrastructure for remote collaboration. Our results, early in the post-Covid era, suggest that the winner-take-all dynamics of economic geography persist \cite{florida2021cities}. 

Some limitations of our study highlight potential future work and extensions. The most obvious extension is to continue collecting data into the future to observe trends as they unfold, for instance to better understand whether OSS activity follows or predicts entrepreneurial activity and innovation. Given the apparent interest in the future of work vis-\'a-vis Covid-19 and remote collaboration, this would be a valuable and important extension. More granular locations of developers, though difficult to source, would allow for studies of clustering of knowledge activity within cities \cite{hegyi2021uid}. The movement of individual developers could provide additional level of detail into how the tech world is adapting to OSS. Movements likely predict future inter-regional linkages, as those who arrive to a new place connect people in their new homes with those from their previous one \cite{kunczer2019benefitting}. Such data also presents the opportunity to study how software innovations diffuse geographically for instance via collaboration ties \cite{toth2021repeated}.

Our approach could be extended to cover alternative platforms for open source contributions including GitLab and Bitbucket, the absence of which may bias our results \cite{trujillo2021penumbra}. There are also potential biases within GitHub itself \cite{kalliamvakou2014promises}, for instance some projects are experiments, class projects, or webpages and not the kind of code that is widely used. Though we think this is unlikely to significantly bias our results, future work using our data should consider this potential limitation. Other possibilities for geolocating developers, including by geocoding the companies and organizations they work for, should be explored to add breadth. However, such extensions may bias results as developers in different countries may link to organizations at different rates. Nor does the use of an email address with a country top-level domain guarantee that an individual in question lives in that country. We have attempted to balance these concerns to provide an accurate measure of OSS activity at various scales.

\subsection*{Policy Implications}
Perhaps the most important next step is to derive policy implications from our results. Though OSS activity is highly clustered now, that does not preclude other regions from increasing their participation in OSS and enjoying the apparent benefits of a strong OSS footprint. After all, regional clusters of creativity are known to rise and fall~\cite{doehne2021long} over long periods of time. We frame these implications along three lines, suggesting potential future work in each direction. First we consider what our results suggest about existing policies promoting OSS. We then consider which kinds of policies more generally, for instance those commonly discussed in the cluster policy literature, are likely to be effective. Finally, we discuss how OSS connects and relates to the rest of the economy.

As discussed in our literature review, national and regional public policy on the use of OSS tends to focus on the potential cost-savings of public sector adoption of OSS. As recent work has shown that local OSS activity has significant positive economic externalities, we suggest that there is ample scope for promoting OSS use in the private sector and at universities. For example, a regional scientific funding agency may encourage or require software outputs of projects to be release under an open source license. Students can be encouraged to contribute to OSS projects in the course of the projects as a way to build and demonstrate expertise. As a sector, OSS is in a unique position to span the branches of the so-called Triple Helix model of innovation policy that connects universities, private sector firms, and public sector institutions \cite{etzkowitz2017triple}. On a more fundamental level, our regression models suggest that human and social development has a strong relationship with OSS activity. From this finding we infer that it is necessary but not sufficient for a place to be wealthy and well-connected to the internet to develop a strong OSS community. Given that we can only explain roughly half of EU regional variance in OSS activity using socio-economic indicators, it seems that there are strong intangible drivers of OSS, suggesting that the right policies can make a difference.

The strong regional clustering observed in our data indicates that insights from the cluster policy literature can inform policymaking on OSS. Effective cluster policy is often more about building networks and sharing information than providing specific economic incentives for an activity \cite{uyarra2016impact}. For example, a city may introduce a mentorship program linking experienced OSS developers with new developers; there is already evidence that one on one or small scale mentorship is one of the best ways to get new people involved in OSS \cite{steinmacher2021being}. On a broader scale, cities and regions can support meetups of open source developers with space and resources. Fostering informal networks can make local ecosystems more resilient \cite{saxenian1990regional} and innovative \cite{trippl2009knowledge}. Firms can be informed about the various benefits of using and contributing to OSS \cite{nagle2018learning}.  On the other hand, policy efforts foster OSS use directly, for instance by mandates for public institutions, have had limited success \cite{blind2021impact}.

Startups and firms that create software could be informed about the potential gains of making their software open source \cite{nagle2018learning,nagle2019open}. For example, a strong open source presence can be an effective signal of competence and ability to potential customers and investors. These kinds of policies represent a significant departure from the most common OSS policies implemented today. Between regions and cities there is also a potential to support networks of developers, not only projects, as the EU already does for traditional R\&D sectors \cite{wanzenbock2020impacts}. Mapping these inter-regional networks of OSS developers would be a valuable next step.

We have also seen that OSS activity relates to innovation outcomes, namely patenting activity and the publication of research papers on AI. While we cannot demonstrate the direction of the effect in this case, i.e. whether OSS activity accelerates innovation activity or is merely a by-product of the process, the correlations we observe suggest this question merits additional study. One potential framework for such an investigation is the concept of relatedness, which suggests that places create new products and inventions by adapting and recombining existing expertise and know-how \cite{essletzbichler2015relatedness,hidalgo2021economic}. From this perspective software expertise may open up specific and valuable new paths to innovation.
Another related idea is the notion of technological complexity, which suggests that innovations that combine elementary ingredients in complex ways are crucial for economic growth at the knowledge frontier
\cite{mewes2020technological,pintar2021complex}. In this context we suggest that software is a special kind of skill that enables people to combine inputs in novel ways \cite{gomez2021estimating}. Understanding software's role in modern innovation, hence its impact on economic growth, is thus an area warranting additional research.

\section*{Acknowledgements}

JW, MN, and AP acknowledge support from the City of Vienna WU Jubilee Fund (Jubil\"aumsfonds der Stadt Wien). The authors thank Frank Neffke, Gerg\"o T\'oth, S\'andor Juh\'asz, and Bal\'azs Lengyel for helpful discussions and Jeremias Br\"andle for assistance with the patent data.

\section*{Conflict of interest}
The authors declare that they have no conflict of interest.

\clearpage
\section*{Appendix}

\begin{table}[!h]
\ra{1} 
\rowcolors{2}{gray!25}{white}
\begin{tabular}{llcrrrrrr}
\toprule
{} &         Country & ISO2 &  \# GitHub &  \# Twitter &  \# Email Suf. &  Total Contributors. &  Pop. (mm) &  Contribs./100k \\
\midrule
1  &         Iceland &   IS &           249 &             10 &                 162 &                       421 &         0.4 &                    105 \\
2  &     Switzerland &   CH &          4978 &             55 &                2164 &                      7197 &         8.6 &                     84 \\
3  &          Norway &   NO &          3137 &             29 &                 846 &                      4012 &         5.3 &                     76 \\
4  &          Sweden &   SE &          6076 &             38 &                1209 &                      7323 &        10.3 &                     71 \\
5  &         Finland &   FI &          3086 &             20 &                 707 &                      3813 &         5.5 &                     69 \\
6  &         Denmark &   DK &          2798 &             22 &                1086 &                      3906 &         5.8 &                     67 \\
7  &     Netherlands &   NL &          8843 &            110 &                1820 &                     10773 &        17.3 &                     62 \\
8  &          Canada &   CA &         19267 &            219 &                2783 &                     22269 &        37.6 &                     59 \\
9  &         Estonia &   EE &           600 &              6 &                 154 &                       760 &         1.3 &                     58 \\
10 &      Luxembourg &   LU &           280 &              4 &                  40 &                       324 &         0.6 &                     54 \\
11 &     New Zealand &   NZ &          2299 &             28 &                 315 &                      2642 &         4.9 &                     54 \\
12 &       Singapore &   SG &          2818 &             33 &                 251 &                      3102 &         5.7 &                     54 \\
13 &         Ireland &   IE &          2224 &             25 &                 282 &                      2531 &         4.9 &                     52 \\
14 &   United States &   US &        128526 &           1831 &               14014 &                    144371 &       328.2 &                     44 \\
15 &  United Kingdom &   GB &         24493 &            443 &                4516 &                     29452 &        66.8 &                     44 \\
16 &       Australia &   AU &          8970 &            120 &                1247 &                     10337 &        25.4 &                     41 \\
17 &         Germany &   DE &         25027 &            276 &                7909 &                     33212 &        83.1 &                     40 \\
18 &         Austria &   AT &          2472 &             53 &                 751 &                      3276 &         8.9 &                     37 \\
19 &          France &   FR &         17474 &            202 &                4875 &                     22551 &        67.1 &                     34 \\
20 &         Belgium &   BE &          3134 &             43 &                 758 &                      3935 &        11.5 &                     34 \\
21 &          Israel &   IL &          2131 &             43 &                 314 &                      2488 &         9.1 &                     27 \\
22 &         Belarus &   BY &          2375 &              8 &                 149 &                      2532 &         9.5 &                     27 \\
23 &        Portugal &   PT &          2485 &             32 &                 285 &                      2802 &        10.3 &                     27 \\
24 &       Lithuania &   LT &           683 &              5 &                  60 &                       748 &         2.8 &                     27 \\
25 &          Poland &   PL &          8864 &             51 &                1491 &                     10406 &        38.0 &                     27 \\
26 &         Czechia &   CZ &          2771 &             34 &                   0 &                      2805 &        10.7 &                     26 \\
27 &        Bulgaria &   BG &          1510 &             11 &                 234 &                      1755 &         7.0 &                     25 \\
28 &        Slovenia &   SI &           411 &              6 &                  75 &                       492 &         2.1 &                     23 \\
29 &          Latvia &   LV &           371 &              4 &                  68 &                       443 &         1.9 &                     23 \\
30 &           Spain &   ES &          9091 &            157 &                1345 &                     10593 &        47.1 &                     22 \\
31 &           Malta &   MT &           100 &              0 &                  12 &                       112 &         0.5 &                     22 \\
32 &          Taiwan &   TW &          4293 &             66 &                 620 &                      4979 &        23.6 &                     21 \\
33 &     South Korea &   KR &         10025 &             35 &                 861 &                     10921 &        51.7 &                     21 \\
34 &         Hungary &   HU &          1616 &             13 &                 184 &                      1813 &         9.8 &                     18 \\
35 &         Croatia &   HR &           666 &              3 &                  73 &                       742 &         4.1 &                     18 \\
36 &          Russia &   RU &         15543 &            108 &                9620 &                     25271 &       144.4 &                     18 \\
37 &       Hong Kong &   HK &          1151 &             13 &                 139 &                      1303 &         7.5 &                     17 \\
38 &         Ukraine &   UA &          6941 &             29 &                 234 &                      7204 &        44.4 &                     16 \\
39 &          Serbia &   RS &           953 &              5 &                  81 &                      1039 &         6.9 &                     15 \\
40 &          Cyprus &   CY &           157 &              4 &                   7 &                       168 &         1.2 &                     14 \\
41 &          Greece &   GR &          1338 &             21 &                 151 &                      1510 &        10.7 &                     14 \\
42 &        Slovakia &   SK &           620 &              6 &                  93 &                       719 &         5.5 &                     13 \\
43 &           Japan &   JP &         12181 &            277 &                3248 &                     15706 &       126.3 &                     12 \\
44 &         Uruguay &   UY &           397 &             11 &                  27 &                       435 &         3.5 &                     12 \\
45 &          Brazil &   BR &         24021 &            299 &                1571 &                     25891 &       211.0 &                     12 \\
46 &      Costa Rica &   CR &           501 &              7 &                  85 &                       593 &         5.0 &                     12 \\
47 &           Italy &   IT &          5728 &            107 &                1369 &                      7204 &        60.3 &                     12 \\
48 &         Romania &   RO &          1820 &             11 &                 148 &                      1979 &        19.4 &                     10 \\
49 &         Namibia &   NA &           250 &              9 &                   1 &                       260 &         2.5 &                     10 \\
50 &       Argentina &   AR &          3864 &             50 &                 418 &                      4332 &        44.9 &                     10 \\
\bottomrule
\end{tabular}
\caption{Countries ranked by number of GitHub contributors (located via GitHub or Twitter location, or email suffix data), per 100k inhabitants. We exclude countries with fewer than 300k inhabitants and Montenegro, because the ``.me'' domain suffix is popular world-wide.}
\label{tab:countries_per_100k}
\end{table}

\begin{table}
\ra{1} 
\rowcolors{2}{gray!25}{white}
\begin{tabular}{lrrr}
\toprule
                                       MSA Name &  Count Contributors &  Population &  Contributors/100k \\
\midrule
                                    Boulder, CO &                 995 &      326196 &                305 \\
             San Jose-Sunnyvale-Santa Clara, CA &                4587 &     1990660 &                230 \\
              San Francisco-Oakland-Hayward, CA &               10702 &     4731803 &                226 \\
                    Seattle-Tacoma-Bellevue, WA &                8830 &     3979845 &                221 \\
                                  Ann Arbor, MI &                 600 &      367601 &                163 \\
                           Champaign-Urbana, IL &                 365 &      226033 &                161 \\
                          Austin-Round Rock, TX &                3370 &     2227083 &                151 \\
                         Durham-Chapel Hill, NC &                 759 &      644367 &                117 \\
            Portland-Vancouver-Hillsboro, OR-WA &                2751 &     2492412 &                110 \\
                            Charlottesville, VA &                 238 &      218615 &                108 \\
                 Boston-Cambridge-Newton, MA-NH &                5221 &     4873019 &                107 \\
                     Santa Cruz-Watsonville, CA &                 249 &      273213 &                 91 \\
                     Denver-Aurora-Lakewood, CO &                2555 &     2967239 &                 86 \\
                                    Madison, WI &                 574 &      664865 &                 86 \\
                                    Raleigh, NC &                1159 &     1390785 &                 83 \\
                             Salt Lake City, UT &                 989 &     1232696 &                 80 \\
                   Lafayette-West Lafayette, IN &                 166 &      233002 &                 71 \\
                                    Trenton, NJ &                 251 &      367430 &                 68 \\
                                Gainesville, FL &                 227 &      329128 &                 68 \\
                  Santa Maria-Santa Barbara, CA &                 291 &      446499 &                 65 \\
          New York-Newark-Jersey City, NY-NJ-PA &               11579 &    19216182 &                 60 \\
                                 Provo-Orem, UT &                 390 &      648252 &                 60 \\
                         San Diego-Carlsbad, CA &                1903 &     3338330 &                 57 \\
                      College Station-Bryan, TX &                 148 &      264728 &                 55 \\
                                 Pittsburgh, PA &                1185 &     2317600 &                 51 \\
                               Fort Collins, CO &                 181 &      356899 &                 50 \\
 Nashville-Davidson--Murfreesboro--Franklin, TN &                 952 &     1934317 &                 49 \\
                Burlington-South Burlington, VT &                 104 &      220411 &                 47 \\
                       Athens-Clarke County, GA &                 101 &      213750 &                 47 \\
              Atlanta-Sandy Springs-Roswell, GA &                2668 &     6020364 &                 44 \\
  San Luis Obispo-Paso Robles-Arroyo Grande, CA &                 124 &      283111 &                 43 \\
   Washington-Arlington-Alexandria, DC-VA-MD-WV &                2698 &     6280487 &                 42 \\
                                     Eugene, OR &                 161 &      382067 &                 42 \\
        Minneapolis-St. Paul-Bloomington, MN-WI &                1554 &     3640043 &                 42 \\
                                 Bellingham, WA &                  96 &      229247 &                 41 \\
             Los Angeles-Long Beach-Anaheim, CA &                5533 &    13214799 &                 41 \\
             Chicago-Naperville-Elgin, IL-IN-WI &                3876 &     9458539 &                 40 \\
                                    Lincoln, NE &                 132 &      336374 &                 39 \\
                                 Boise City, ID &                 272 &      749202 &                 36 \\
                                  Rochester, NY &                 376 &     1069644 &                 35 \\
                                     Tucson, AZ &                 347 &     1047279 &                 33 \\
                                 Santa Rosa, CA &                 154 &      494336 &                 31 \\
                  Orlando-Kissimmee-Sanford, FL &                 804 &     2608147 &                 30 \\
    Philadelphia-Camden-Wilmington, PA-NJ-DE-MD &                1883 &     6102434 &                 30 \\
                          Vallejo-Fairfield, CA &                 138 &      447643 &                 30 \\
              Charlotte-Concord-Gastonia, NC-SC &                 771 &     2636883 &                 29 \\
                             Kansas City, MO-KS &                 637 &     2157990 &                 29 \\
                                  Rochester, MN &                  66 &      221921 &                 29 \\
                                   Columbus, OH &                 603 &     2122271 &                 28 \\
                                   Fargo, ND-MN &                  67 &      246145 &                 27 \\
\bottomrule
\end{tabular}
\caption{Top US MSAs with population of at least 250k, by developers per capita.}
\label{tab:msa_50}
\end{table}

\rowcolors{2}{gray!25}{white}
\begin{table}[!h]
\centering
\ra{1.1} 
\begin{tabular}{lccc}
\toprule
 Country &  $AGC_{TertEdu}$ &  $AGC_{HiTech}$ &  $AGC_{OSS}$ \\
\midrule
     PT &             0.10 &            0.15 &         0.22 \\
     IT &             0.07 &            0.19 &         0.22 \\
     FR &             0.08 &            0.19 &         0.28 \\
     BE &             0.06 &            0.12 &         0.29 \\
     EL &             0.09 &            0.31 &         0.29 \\
     SE &             0.05 &            0.20 &         0.30 \\
     ES &             0.07 &            0.24 &         0.31 \\
     IE &             0.05 &            0.09 &         0.31 \\
     HR &             0.02 &            0.25 &         0.32 \\
     NL &             0.06 &            0.10 &         0.32 \\
     CH &             0.05 &            0.10 &         0.33 \\
     FI &             0.05 &            0.17 &         0.36 \\
     DK &             0.11 &            0.28 &         0.38 \\
     AT &             0.06 &            0.17 &         0.39 \\
     UK &             0.08 &            0.17 &         0.40 \\
     PL &             0.08 &            0.22 &         0.41 \\
     DE &             0.06 &            0.16 &         0.42 \\
     NO &             0.06 &            0.24 &         0.44 \\
     RO &             0.17 &            0.38 &         0.47 \\
     BG &             0.17 &            0.39 &         0.54 \\
     SK &             0.10 &            0.15 &         0.59 \\
     RS &             0.19 &            0.40 &         0.61 \\
     SI &             0.14 &            0.23 &         0.61 \\
     LT &             0.14 &            0.52 &         0.64 \\
     HU &             0.18 &            0.26 &         0.66 \\
     CZ &             0.14 &            0.18 &         0.67 \\
\bottomrule
\end{tabular}
\caption{The concentration of workers with higher education, workers in high-tech industries, and OSS contributors across NUTS2 regions of countries, measured by the Adjusted Geographic Concentration (AGC). The concentration of OSS contributors in particular regions is consistently higher than the concentration of both alternative populations.}
\label{tab:nuts_agc}
\end{table}

\subsection*{Summary Statistics and Robustness Tests}
In this section we report numerous summary statistics about our data as well as robustness tests to supplement our regression analyses.

\begin{table}[!h]
\begin{tabular}{lrrrrrrrr}
\toprule
{} & OSS /mm &  PPP GNI PC&  HDI 2019 &  WVS Trust &  IPI 2019 &    ECI &  AI/DL papers &  Internet Pen. \\
\midrule
N Obs. &     183 &        176 &    178 &          75 &   117 &  157 &                183 &          180 \\
Mean  &       3.27 &         21.52 &      0.72 &          25.05 &     6.62 &   -0.00 &                 -4.63 &           54.95 \\
Stdev.   &       1.99 &         21.18 &      0.15 &          18.35 &     1.54 &    1.00 &                  4.92 &           29.04 \\
Min.   &      -1.20 &          0.79 &      0.39 &           2.10 &     2.35 &   -2.32 &                 -9.21 &            2.00 \\
25\%   &       1.95 &          5.18 &      0.60 &          12.00 &     5.79 &   -0.73 &                 -9.21 &           27.88 \\
50\%   &       3.34 &         13.64 &      0.74 &          20.60 &     6.52 &   -0.05 &                 -9.21 &           58.50 \\
75\%   &       4.85 &         32.63 &      0.84 &          33.30 &     7.85 &    0.70 &                  0.11 &           80.12 \\
Max   &       7.54 &         92.27 &      0.96 &          73.90 &     9.61 &    2.27 &                  4.22 &           98.30 \\
\bottomrule
\end{tabular}
\caption{Summary statistics table of key national statistics.}
\label{tab:national_sumstat}
\end{table}

\begin{landscape}

\begin{table}[h!]
\begin{tabular}{lrrrrrrrrrrr}
\toprule
{} &  OSS &  GDP PC &  Pop &  PopDens &  EVSTrust &   EQI &  R\&D Spd &  \%HighTech &  EEPatents &  Patents&  \%TertEdu \\
\midrule
N Obs. &                297 &      285 &  297 &             288 &     218 &  149 &                 259 &                279 &                                      259 &             259 &              293 \\
Mean  &                  1.00 &        4.40 &    6.13 &               4.97 &       0.38 &   -0.18 &                   3.10 &                  4.11 &                                       -0.14 &               1.47 &               33.93 \\
Stdev.   &                  0.49 &        0.25 &    0.35 &               1.20 &       0.20 &    0.99 &                   1.32 &                  2.25 &                                        1.41 &               1.57 &               10.17 \\
Min   &                 -0.79 &        3.61 &    4.48 &               1.19 &       0.00 &   -2.26 &                  -1.45 &                  0.70 &                                       -2.30 &              -2.70 &               11.80 \\
25\%   &                  0.71 &        4.26 &    5.94 &               4.29 &       0.22 &   -0.95 &                   2.29 &                  2.50 &                                       -1.26 &               0.28 &               26.40 \\
50\%   &                  1.03 &        4.46 &    6.16 &               4.86 &       0.33 &   -0.29 &                   3.14 &                  3.70 &                                       -0.18 &               1.76 &               33.10 \\
75\%   &                  1.32 &        4.58 &    6.34 &               5.66 &       0.53 &    0.50 &                   4.09 &                  5.15 &                                        0.90 &               2.64 &               41.10 \\
Max   &                  2.24 &        4.98 &    7.09 &               8.91 &       0.81 &    2.32 &                   6.48 &                 12.90 &                                        3.37 &               4.53 &               59.86 \\
\bottomrule
\end{tabular}
\caption{Summary statistics table of key NUTS2 statistics.}
\label{tab:nuts_sumstat}
\end{table}

\begin{table}[h!]
\begin{tabular}{lrrrrrrrr}
\toprule
{} &  OSS /mm &  PPP GNI PC&  HDI 2019 &  WVS Trust &  IPI 2019 &    ECI &  AI/DL papers &  Internet Pen. \\
\midrule
OSS /mm            &      1.000 &         0.785 &     0.860 &          0.677 &    0.874 &  0.820 &                 0.665 &           0.775 \\
PPP GNI PC        &      0.785 &         1.000 &     0.963 &          0.698 &    0.846 &  0.823 &                 0.710 &           0.909 \\
HDI 2019             &      0.860 &         0.963 &     1.000 &          0.681 &    0.866 &  0.847 &                 0.743 &           0.906 \\
WVS Trust        &      0.677 &         0.698 &     0.681 &          1.000 &    0.659 &  0.619 &                 0.602 &           0.642 \\
IPI 2019              &      0.874 &         0.846 &     0.866 &          0.659 &    1.000 &  0.798 &                 0.734 &           0.763 \\
ECI                  &      0.820 &         0.823 &     0.847 &          0.619 &    0.798 &  1.000 &                 0.757 &           0.831 \\
AI/DL papers &      0.665 &         0.710 &     0.743 &          0.602 &    0.734 &  0.757 &                 1.000 &           0.712 \\
Internet Pen.       &      0.775 &         0.909 &     0.906 &          0.642 &    0.763 &  0.831 &                 0.712 &           1.000 \\
\bottomrule
\end{tabular}
\caption{Spearman Correlation table of key national statistics.}
\label{tab:national_corr}
\end{table}

\begin{table}[!h]
\begin{tabular}{lrrrrrrrrrrr}
\toprule
{} &  OSS &  GDP PC &  Pop &  PopDens &  EVSTrust &   EQI &  R\&D Spd &  \%HighTech &  EEPatents &  Patents&  \%TertEdu \\\\
\midrule
OSS                       &                 1.000 &       0.567 &   0.249 &              0.355 &      0.471 &  0.455 &                  0.368 &                 0.605 &                                       0.542 &              0.542 &               0.664 \\
GDP PC                                 &                 0.567 &       1.000 &   0.025 &              0.321 &      0.675 &  0.690 &                  0.791 &                 0.441 &                                       0.743 &              0.837 &               0.579 \\
Pop                                     &                 0.249 &       0.025 &   1.000 &              0.406 &     -0.282 & -0.202 &                 -0.414 &                 0.235 &                                       0.280 &              0.192 &               0.045 \\
PopDens                         &                 0.355 &       0.321 &   0.406 &              1.000 &     -0.100 &  0.115 &                  0.130 &                 0.457 &                                       0.269 &              0.300 &               0.258 \\
EVSTrust                                  &                 0.471 &       0.675 &  -0.282 &             -0.100 &      1.000 &  0.658 &                  0.655 &                 0.253 &                                       0.528 &              0.588 &               0.546 \\
EQI                                      &                 0.455 &       0.690 &  -0.202 &              0.115 &      0.658 &  1.000 &                  0.672 &                 0.177 &                                       0.579 &              0.711 &               0.662 \\
R\&D Spd                      &                 0.368 &       0.791 &  -0.414 &              0.130 &      0.655 &  0.672 &                  1.000 &                 0.407 &                                       0.568 &              0.671 &               0.495 \\
\%HighTech                       &                 0.605 &       0.441 &   0.235 &              0.457 &      0.253 &  0.177 &                  0.407 &                 1.000 &                                       0.505 &              0.450 &               0.586 \\
EEPatents &                 0.542 &       0.743 &   0.280 &              0.269 &      0.528 &  0.579 &                  0.568 &                 0.505 &                                       1.000 &              0.886 &               0.442 \\
Patents                        &                 0.542 &       0.837 &   0.192 &              0.300 &      0.588 &  0.711 &                  0.671 &                 0.450 &                                       0.886 &              1.000 &               0.455 \\
\%TertEdu                         &                 0.664 &       0.579 &   0.045 &              0.258 &      0.546 &  0.662 &                  0.495 &                 0.586 &                                       0.442 &              0.455 &               1.000 \\
\bottomrule
\end{tabular}
\caption{Spearman Correlation table of NUTS2 statistics.}
\label{tab:nuts_corr}
\end{table}

\end{landscape}

\begin{table}[!h]
\begin{tabular}{llllll}
\toprule
{} &    M1 &     M2 &     M3 &     M4 &     M5 \\
\midrule
PPP GNI per Cap.  &   2.7 &    3.2 &    3.8 &    3.2 &    3.5 \\
Internet Penetration &   2.7 &    5.8 &    3.3 &    3.7 &    6.6 \\
Population        &     1 &    1.1 &    1.1 &    1.1 &    1.2 \\
HDI      &   N/A &    6.8 &    N/A &    N/A &    8.3 \\
IPI       &   N/A &    N/A &    3.4 &    N/A &    N/A \\
ECI            &   N/A &    N/A &    N/A &    3.4 &    3.7 \\
\bottomrule
\end{tabular}
\caption{Variance Inflation Factors (VIF) for country-level regression models, models 1-5. Despite significant pairwise correlations (see correlation table), VIF scores remain at reasonable levels.}
\label{tab:national_vif}
\end{table}

\begin{table}[!h]
\begin{tabular}{llllllll}
\toprule
{} &     M1 &     M2 &     M3 &     M4 &      M5 &      M6 &      M7 \\
\midrule
Internet Pen.                              &    2.2 &    2.8 &        2.3 &     2.3 &     2.4 &     2.5 &    2.3 \\
GDP PC                                &    2.3 &    2.8 &     6.2 &     2.6 &     3.8 &     4.5 &    2.5 \\
Pop                                     &    1.2 &    1.3 &      2.6 &     1.2 &     1.5 &     1.4 &    1.2 \\
PopDens                          &    1.4 &    1.5 &     1.4 &     1.5 &     1.5 &     1.5 &    1.4 \\
EVSTrust                                  &    N/A &    2.4  &    N/A &     N/A &     N/A &     N/A &    N/A \\

R\&D Spd                      &    N/A &    N/A &      6.3 &     N/A &     N/A &     N/A &    N/A \\
\%HighTech                      &    N/A &      N/A &    N/A &     1.5 &     N/A &     N/A &    N/A \\
EEPatents &    N/A &    N/A &    N/A &       N/A &     2.8 &     N/A &    N/A \\
Patents                          &    N/A &     N/A &    N/A &     N/A &     N/A &     4.2 &    N/A \\
\%TertEdu                         &    N/A &      N/A &    N/A &     N/A &     N/A &     N/A &    1.5 \\
\bottomrule
\end{tabular}
\caption{Variance Inflation Factors (VIF) for NUTS2-level regression models. Despite significant pairwise correlations (see correlation table), VIF scores remain at reasonable levels.}
\label{tab:national_vif}
\end{table}

\subsection*{Spatial Regression Diagnostics}
In this section we report linear regression models fit via OLS on the NUTS2 data in Table \ref{tab:ols_nuts_regressions}, as well as their spatial regression diagnostics. As suggested in the main text, we observe significant spatial autocorrelation of the key dependent variable (OSS contributors per 100k inhabitants) via Moran's I. We also examined spatial autocorrelation among the residuals of linear regression models fit via OLS using Moran's I and a Lagrange Multiplier test \cite{cliff1981spatial,anselin1988lagrange}. We find significant spatial autocorrelation, motivating the use of the GMM model \cite{arraiz2010spatial} in the manuscript. We note that despite these (necessary) adjustments, the results (in terms of estimated effect sizes and statistical significance) are highly similar.

\begin{table}[] \centering
\begin{tabular}{@{\extracolsep{5pt}}lccccccc}
\\[-1.8ex]\hline
& \multicolumn{7}{c}{Active GitHub Contributors/100k Inhab. European NUTS2 (log, 2021)} \
\cr
\\[-1.8ex] & (1) & (2) & (3) & (4) & (5) & (6) & (7) \\
\hline \\[-1.8ex]
  Internet Penetration & 0.010$^{**}$ & 0.007$^{}$ & 0.012$^{**}$ & 0.011$^{***}$ & 0.011$^{**}$ & 0.011$^{**}$ & 0.004$^{}$ \\
  ---(\% of Pop. 2017)  & (0.005) & (0.006) & (0.005) & (0.004) & (0.004) & (0.004) & (0.004) \\
 GDP per Cap. & 0.754$^{***}$ & 0.694$^{***}$ & 0.014$^{}$ & 0.450$^{***}$ & 0.274$^{}$ & 0.342$^{}$ & 0.426$^{***}$ \\
 ---(log Eur, 2017)   & (0.147) & (0.167) & (0.262) & (0.152) & (0.181) & (0.211) & (0.155) \\
 Population  & 0.270$^{***}$ & 0.381$^{***}$ & 0.486$^{***}$ & 0.266$^{***}$ & 0.169$^{**}$ & 0.216$^{**}$ & 0.285$^{***}$ \\
 ---(log, 2020)   & (0.083) & (0.095) & (0.136) & (0.087) & (0.085) & (0.085) & (0.079) \\
 Population Dens. & 0.043$^{*}$ & 0.052$^{**}$ & 0.048$^{}$ & -0.019$^{}$ & 0.064$^{**}$ & 0.058$^{*}$ & 0.027$^{}$ \\
 ---(log, 2017) & (0.026) & (0.026) & (0.029) & (0.023) & (0.029) & (0.030) & (0.023) \\
  EVS Trust  & & 0.415$^{**}$ & & & & & \\
 ---(2017) & & (0.176) & & & & & \\
 R\&D Spend. per Cap. & & & 0.128$^{**}$ & & & & \\
 ---(log, 2017)  & & & (0.052) & & & & \\
 \% Empl. High-Tech & & & & 0.089$^{***}$ & & & \\
 ---(2019/20) & & & & (0.011) & & & \\
 Patents Elec-Eng./100k & & & & & 0.072$^{***}$ & & \\
 ---(log, 2017)  & & & & & (0.023) & & \\
 Patents/100k  & & & & & & 0.050$^{*}$ & \\
 ---(log, 2017)  & & & & & & (0.027) & \\
 \% with Tertiary Edu. & & & & & & & 0.022$^{***}$ \\
 ---(2019/20) & & & & & & & (0.002) \\
\hline \\[-1.8ex]
 Observations & 276 & 198 & 258 & 262 & 258 & 258 & 276 \\
 Adjusted $R^2$ & 0.388 & 0.428 & 0.410 & 0.503 & 0.406 & 0.395 & 0.530 \\
 Residual Std. Error & 0.371 & 0.354 & 0.347 & 0.328 & 0.348 & 0.352 & 0.325  \\
 F Statistic & 36.8$^{***}$  & 27.1$^{***}$  & 34.5$^{***}$  & 50.0$^{***}$  & 33.0$^{***}$  & 31.0$^{***}$  & 52.8$^{***}$ 
 \\
\hline
& \multicolumn{7}{r}{$^{*}$p$<$0.1; $^{**}$p$<$0.05; $^{***}$p$<$0.01} \\
\end{tabular}
\caption{OLS regression models relating EU NUTS2 counts of GitHub contributors per 100k inhabitants (log-transformed) and socio-economic indicators. We report (generic) heteroskedasticity robust standard errors. The models in this table do not consider the potential bias introduced by spatial autocorrelation. The model diagnostics in the following table suggest that these results may not be robust.}
\label{tab:ols_nuts_regressions}
\end{table}

\begin{table}
\begin{tabular}{lrrrrrrr}
\toprule
{} &  Model 1 &  Model 2 &  Model 3 &  Model 4 &  Model 5 &  Model 6 &  Model 7 \\
\midrule
Moran's I (residual)          &     3.16 &     3.36 &     2.03 &     2.67 &     1.85 &     2.36 &     1.89 \\
Moran's I p-value             &     0.00 &     0.00 &     0.04 &     0.01 &     0.06 &     0.02 &     0.06 \\
Lagrange Multiple error       &     8.91 &     9.82 &     3.46 &     6.10 &     2.87 &     4.80 &     3.00 \\
Lagrange Multiple error prob. &     0.00 &     0.00 &     0.06 &     0.01 &     0.09 &     0.03 &     0.08 \\
\bottomrule
\end{tabular}
\caption{OLS regression spatial diagnostics for the models in Table \ref{tab:ols_nuts_regressions} relating EU NUTS2 counts of GitHub contributors per 100k inhabitants (log-transformed) and socio-economic indicators. These tests indicate significant spatial correlations, necessitating the use of the more sophisticated GMM model in the text to adjust for potential biases.}
\label{tab:ols_nuts_regressions}
\end{table}


\clearpage

\setstretch{1.0}

\bibliographystyle{spmpsci}       % APS-like style for physics
\bibliography{bibliography.bib}   % name your BibTeX data base

\begin{thebibliography}{10}
\providecommand{\url}[1]{{#1}}
\providecommand{\urlprefix}{URL }
\expandafter\ifx\csname urlstyle\endcsname\relax
  \providecommand{\doi}[1]{DOI~\discretionary{}{}{}#1}\else
  \providecommand{\doi}{DOI~\discretionary{}{}{}\begingroup
  \urlstyle{rm}\Url}\fi

\bibitem{aksulu2010comprehensive}
Aksulu, A., Wade, M.R.: A comprehensive review and synthesis of open source
  research.
\newblock Journal of the Association for Information Systems \textbf{11}(11), 6
  (2010)

\bibitem{albusays2021diversity}
Albusays, K., Bjorn, P., Dabbish, L., Ford, D., Murphy-Hill, E., Serebrenik,
  A., Storey, M.A.: The diversity crisis in software development.
\newblock IEEE Software \textbf{38}(2), 19--25 (2021)

\bibitem{alhusen2021new}
Alhusen, H., Bennat, T., Bizer, K., Cantner, U., Horstmann, E., Kalthaus, M.,
  Proeger, T., Sternberg, R., T{\"o}pfer, S.: {A new measurement conception for
  the ‘doing-using-interacting’ mode of innovation}.
\newblock Res. Pol. \textbf{50}(4) (2021)

\bibitem{andreessen2011software}
Andreessen, M.: Why software is eating the world.
\newblock Wall Street Journal \textbf{20}(2011), C2 (2011)

\bibitem{anselin1988lagrange}
Anselin, L.: Lagrange multiplier test diagnostics for spatial dependence and
  spatial heterogeneity.
\newblock Geographical analysis \textbf{20}(1), 1--17 (1988)

\bibitem{anthes2016open}
Anthes, G.: Open source software no longer optional.
\newblock Communications of the ACM \textbf{59}(8), 15--17 (2016)

\bibitem{arraiz2010spatial}
Arraiz, I., Drukker, D.M., Kelejian, H.H., Prucha, I.R.: A spatial
  cliff-ord-type model with heteroskedastic innovations: Small and large sample
  results.
\newblock Journal of Regional Science \textbf{50}(2), 592--614 (2010)

\bibitem{balland2021mapping}
Balland, P.A., Boschma, R.: Mapping the potentials of regions in europe to
  contribute to new knowledge production in industry 4.0 technologies.
\newblock Regional Studies pp. 1--15 (2021)

\bibitem{bettencourt2002client}
Bettencourt, L.A., Ostrom, A.L., Brown, S.W., Roundtree, R.I.: Client
  co-production in knowledge-intensive business services.
\newblock California management review \textbf{44}(4), 100--128 (2002)

\bibitem{blind2021impact}
Blind, K., B{\"o}hm, M., Grzegorzweska, P., Katz, A., Muto, S., P{\"a}tsch, S.,
  Schubert, T.: The impact of open source software and hardware on
  technological independence, competitiveness and innovation in the eu economy
  (2021)

\bibitem{boschma2005proximity}
Boschma, R.: Proximity and innovation: a critical assessment.
\newblock Regional studies \textbf{39}(1), 61--74 (2005)

\bibitem{boschma2007knowledge}
Boschma, R.A., Ter~Wal, A.L.: Knowledge networks and innovative performance in
  an industrial district: the case of a footwear district in the south of
  italy.
\newblock Industry and innovation \textbf{14}(2), 177--199 (2007)

\bibitem{branstetter2019get}
Branstetter, L.G., Drev, M., Kwon, N.: Get with the program: Software-driven
  innovation in traditional manufacturing.
\newblock Management Science \textbf{65}(2), 541--558 (2019)

\bibitem{cliff1981spatial}
Cliff, A.D., Ord, J.K.: Spatial processes: models \& applications.
\newblock Taylor \& Francis (1981)

\bibitem{dahlander2021open}
Dahlander, L., Gann, D.M., Wallin, M.W.: How open is innovation? a
  retrospective and ideas forward.
\newblock Research Policy \textbf{50}(4), 104218 (2021)

\bibitem{dahlander2006man}
Dahlander, L., Wallin, M.W.: A man on the inside: Unlocking communities as
  complementary assets.
\newblock Research policy \textbf{35}(8), 1243--1259 (2006)

\bibitem{dahlberg2021quality}
Dahlberg, S., Sundstr{\"o}m, A., Holmberg, S., Rothstein, B., Alvarado~Pachon,
  N., Dalli, C.M.: {The Quality of Government Basic Dataset}.
\newblock University of Gothenburg: The Quality of Government Institute  (2021)

\bibitem{doehne2021long}
Doehne, M., Rost, K.: Long waves in the geography of innovation: The rise and
  decline of regional clusters of creativity over time.
\newblock Research Policy \textbf{50}(9), 104298 (2021)

\bibitem{eghbal2020working}
Eghbal, N.: Working in public: the making and maintenance of open source
  software.
\newblock Stripe Press (2020)

\bibitem{ellison1997geographic}
Ellison, G., Glaeser, E.L.: Geographic concentration in us manufacturing
  industries: a dartboard approach.
\newblock Journal of political economy \textbf{105}(5), 889--927 (1997)

\bibitem{essletzbichler2015relatedness}
Essletzbichler, J.: Relatedness, industrial branching and technological
  cohesion in us metropolitan areas.
\newblock Regional Studies \textbf{49}(5), 752--766 (2015)

\bibitem{etzkowitz2017triple}
Etzkowitz, H., Zhou, C.: The triple helix: University--industry--government
  innovation and entrepreneurship.
\newblock Routledge (2017)

\bibitem{fackler2020gravity}
Fackler, T., Laurentsyeva, N.: Gravity in online collaborations: Evidence from
  github.
\newblock In: CESifo Forum, vol.~21, pp. 15--20. M{\"u}nchen: ifo
  Institut-Leibniz-Institut f{\"u}r Wirtschaftsforschung an der~… (2020)

\bibitem{fershtman2011direct}
Fershtman, C., Gandal, N.: Direct and indirect knowledge spillovers: the
  “social network” of open-source projects.
\newblock The RAND Journal of Economics \textbf{42}(1), 70--91 (2011)

\bibitem{florida2021cities}
Florida, R., Rodr{\'\i}guez-Pose, A., Storper, M.: Cities in a post-covid
  world.
\newblock Urban Studies  (2021)

\bibitem{forman2019digital}
Forman, C., van Zeebroeck, N.: Digital technology adoption and knowledge flows
  within firms: Can the internet overcome geographic and technological
  distance?
\newblock Research Policy \textbf{48}(8), 103697 (2019)

\bibitem{frenken2015industrial}
Frenken, K., Cefis, E., Stam, E.: Industrial dynamics and clusters: a survey.
\newblock Regional studies \textbf{49}(1), 10--27 (2015)

\bibitem{fry2020dataset}
Fry, T., Dey, T., Karnauch, A., Mockus, A.: A dataset and an approach for
  identity resolution of 38 million author ids extracted from 2b git commits.
\newblock In: Proc. of the 17th Int. Conf. on Mining Software Repositories
  (2020)

\bibitem{gerosa2021shifting}
Gerosa, M., Wiese, I., Trinkenreich, B., Link, G., Robles, G., Treude, C.,
  Steinmacher, I., Sarma, A.: The shifting sands of motivation: Revisiting what
  drives contributors in open source.
\newblock In: 2021 IEEE/ACM 43rd International Conference on Software
  Engineering (ICSE) (2021)

\bibitem{EVS2017}
{GESIS Data Archive}: European values study 2017: Integrated dataset (2017).
\newblock \doi{10.4232/1.13560}

\bibitem{gomez2021estimating}
Gomez-Lievano, A., Patterson-Lomba, O.: Estimating the drivers of urban
  economic complexity and their connection to economic performance.
\newblock Royal Society Open Science \textbf{8}(9), 210670 (2021)

\bibitem{gonzalez2008geographic}
Gonzalez-Barahona, J.M., Robles, G., Andradas-Izquierdo, R., Ghosh, R.A.:
  Geographic origin of libre software developers.
\newblock Information Economics and Policy \textbf{20}(4), 356--363 (2008)

\bibitem{gousios2012ghtorrent}
Gousios, G., Spinellis, D.: Ghtorrent: Github's data from a firehose.
\newblock In: 2012 9th IEEE Working Conference on Mining Software Repositories
  (MSR), pp. 12--21. IEEE (2012)

\bibitem{greenstein2014digital}
Greenstein, S., Nagle, F.: {Digital dark matter and the economic contribution
  of Apache}.
\newblock Res. Pol. \textbf{43}(4) (2014)

\bibitem{grier2015tyranny}
Grier, D.A.: The tyranny of geography.
\newblock IEEE Annals of the History of Computing \textbf{48}(02), 100--100
  (2015)

\bibitem{haerpferm}
Haerpfer, C., Inglehart, R., Moreno, A., Welzel, C., Kizilova, K.,
  Diez-Medrano, J.: Round seven-country-pooled datafile.
\newblock World Values Survey  (2017)

\bibitem{hdi2019human}
{HDI}: {Human Development Index}.
\newblock United Nation Development Program  (2019)

\bibitem{hecht2011tweets}
Hecht, B., Hong, L., Suh, B., Chi, E.H.: Tweets from justin bieber's heart: the
  dynamics of the location field in user profiles.
\newblock In: {Proc. of the SIGCHI conference on Human Factors in Computing
  Systems}, pp. 237--246 (2011)

\bibitem{hegyi2021uid}
Hegyi, F., Zhu, M., Janosov, M.: Measuring the impact of urban innovation
  districts.
\newblock Publications Office of the European Union  (2021).
\newblock \doi{10.2760/11053}

\bibitem{hidalgo2021economic}
Hidalgo, C.A.: Economic complexity theory and applications.
\newblock Nature Reviews Physics pp. 1--22 (2021)

\bibitem{hu2021bioinformatics}
Hu, T., Li, J., Zhou, H., Li, C., Holmes, E.C., Shi, W.: Bioinformatics
  resources for sars-cov-2 discovery and surveillance.
\newblock Briefings in bioinformatics  (2021)

\bibitem{jensen2007forms}
Jensen, M.B., Johnson, B., Lorenz, E., Lundvall, B.{\AA}., Lundvall, B.: Forms
  of knowledge and modes of innovation.
\newblock The learning economy and the economics of hope \textbf{155} (2007)

\bibitem{juhasz2021spinoffs}
Juh{\'a}sz, S.: Spinoffs and tie formation in cluster knowledge networks.
\newblock Small Business Economics \textbf{56}(4), 1385--1404 (2021)

\bibitem{juhasz2018creation}
Juh{\'a}sz, S., Lengyel, B.: Creation and persistence of ties in cluster
  knowledge networks.
\newblock Journal of Economic Geography \textbf{18}(6), 1203--1226 (2018)

\bibitem{kalliamvakou2014promises}
Kalliamvakou, E., Gousios, G., Blincoe, K., Singer, L., German, D.M., Damian,
  D.: The promises and perils of mining github.
\newblock In: Proceedings of the 11th working conference on mining software
  repositories, pp. 92--101 (2014)

\bibitem{kaminski2019new}
Kaminski, J., Hopp, C., Tykvov{\'a}, T.: New technology assessment in
  entrepreneurial financing--does crowdfunding predict venture capital
  investments?
\newblock Technological Forecasting and Social Change \textbf{139}, 287--302
  (2019)

\bibitem{klepper2010origin}
Klepper, S.: The origin and growth of industry clusters: The making of silicon
  valley and detroit.
\newblock Journal of Urban Economics \textbf{67}(1), 15--32 (2010)

\bibitem{klinger2021deep}
Klinger, J., Mateos-Garcia, J., Stathoulopoulos, K.: Deep learning, deep
  change? mapping the evolution and geography of a general purpose technology.
\newblock Scientometrics pp. 1--33 (2021)

\bibitem{krugman1991increasing}
Krugman, P.: Increasing returns and economic geography.
\newblock Journal of political economy \textbf{99}(3), 483--499 (1991)

\bibitem{kunczer2019benefitting}
Kunczer, V., Lindner, T., Puck, J.: Benefitting from immigration: The value of
  immigrants’ country knowledge for firm internationalization.
\newblock Journal of International Business Policy \textbf{2}(4), 356--375
  (2019)

\bibitem{lakhani2004open}
Lakhani, K.R., Von~Hippel, E.: How open source software works:“free”
  user-to-user assistance.
\newblock In: Produktentwicklung mit virtuellen Communities, pp. 303--339.
  Springer (2004)

\bibitem{lee2006government}
Lee, J.A.: Government policy toward open source software: The puzzles of
  neutrality and competition.
\newblock Knowledge, Technology \& Policy \textbf{18}(4), 113--141 (2006)

\bibitem{lerner2002some}
Lerner, J., Tirole, J.: Some simple economics of open source.
\newblock The journal of industrial economics \textbf{50}(2), 197--234 (2002)

\bibitem{lima2014coding}
Lima, A., Rossi, L., Musolesi, M.: Coding together at scale: Github as a
  collaborative social network.
\newblock In: Eighth international AAAI conference on weblogs and social media
  (2014)

\bibitem{maraut2008oecd}
Maraut, S., Dernis, H., Webb, C., Spiezia, V., Guellec, D.: The oecd regpat
  database: a presentation  (2008)

\bibitem{may2019gender}
May, A., Wachs, J., Hann{\'a}k, A.: Gender differences in participation and
  reward on stack overflow.
\newblock Empirical Software Engineering \textbf{24}(4), 1997--2019 (2019)

\bibitem{mewes2020technological}
Mewes, L., Broekel, T.: Technological complexity and economic growth of
  regions.
\newblock Research Policy p. 104156 (2020)

\bibitem{montandon2019identifying}
Montandon, J.E., Silva, L.L., Valente, M.T.: {Identifying experts in software
  libraries and frameworks among GitHub users}.
\newblock In: 2019 IEEE/ACM 16th Int. Conf. on Mining Software Repositories
  (MSR) (2019)

\bibitem{mungiu2016measuring}
Mungiu-Pippidi, A., Dada{\v{s}}ov, R.: Measuring control of corruption by a new
  index of public integrity.
\newblock European Journal on Criminal Policy and Research \textbf{22}(3),
  415--438 (2016)

\bibitem{nagle2018learning}
Nagle, F.: Learning by contributing: Gaining competitive advantage through
  contribution to crowdsourced public goods.
\newblock Organization Science \textbf{29}(4), 569--587 (2018)

\bibitem{nagle2019government}
Nagle, F.: Government technology policy, social value, and national
  competitiveness.
\newblock Harvard Business School Strategy Unit Working Paper (19-103) (2019)

\bibitem{nagle2019open}
Nagle, F.: Open source software and firm productivity.
\newblock Management Science \textbf{65}(3), 1191--1215 (2019)

\bibitem{nagle2020report}
Nagle, F., Wheeler, D., Lifshitz-Assaf, H., Ham, H., Hoffman, J.: Report on the
  2020 foss contributor survey.
\newblock The Linux Foundation Core Infrastructure Initiative  (2020)

\bibitem{neffke2019value}
Neffke, F.M.: The value of complementary co-workers.
\newblock Science advances \textbf{5}(12), eaax3370 (2019)

\bibitem{overney2020not}
Overney, C., Meinicke, J., K{\"a}stner, C., Vasilescu, B.: How to not get rich:
  An empirical study of donations in open source.
\newblock In: Proceedings of the ACM/IEEE 42nd International Conference on
  Software Engineering, pp. 1209--1221 (2020)

\bibitem{pietri2019software}
Pietri, A., Spinellis, D., Zacchiroli, S.: {The Software Heritage graph
  dataset: public software development under one roof}.
\newblock In: 2019 IEEE/ACM 16th Int. Conf. on Mining Software Repositories,
  pp. 138--142. IEEE (2019)

\bibitem{pintar2021complex}
Pintar, N., Scherngell, T.: The complex nature of regional knowledge
  production: Evidence on european regions.
\newblock Res. Pol.  (2021)

\bibitem{prana2021including}
Prana, G.A.A., Ford, D., Rastogi, A., Lo, D., Purandare, R., Nagappan, N.:
  Including everyone, everywhere: Understanding opportunities and challenges of
  geographic gender-inclusion in oss.
\newblock IEEE Trans. Soft. Eng.  (2021)

\bibitem{raymond1999cathedral}
Raymond, E.: The cathedral and the bazaar.
\newblock Knowledge, Technology \& Policy \textbf{12}(3), 23--49 (1999)

\bibitem{rey2010pysal}
Rey, S.J., Anselin, L.: Pysal: A python library of spatial analytical methods.
\newblock In: Handbook of applied spatial analysis, pp. 175--193. Springer
  (2010)

\bibitem{riedl2018learning}
Riedl, C., Seidel, V.P.: Learning from mixed signals in online innovation
  communities.
\newblock Organization Science \textbf{29}(6), 1010--1032 (2018)

\bibitem{rothstein2005all}
Rothstein, B., Uslaner, E.M.: All for all: Equality, corruption, and social
  trust.
\newblock World Pol. \textbf{58}, 41 (2005)

\bibitem{rovolis2006ethnic}
Rovolis, A., Tragaki, A.: Ethnic characteristics and geographical distribution
  of immigrants in greece.
\newblock European Urban and Regional Studies \textbf{13}(2), 99--111 (2006)

\bibitem{sahay2019free}
Sahay, S.: Free and open source software as global public goods? what are the
  distortions and how do we address them?
\newblock The Electronic Journal of Information Systems in Developing Countries
  \textbf{85}(4), e12080 (2019)

\bibitem{saxenian1990regional}
Saxenian, A.: Regional networks and the resurgence of silicon valley.
\newblock {California Management Review} \textbf{33}(1), 89--112 (1990)

\bibitem{secundo2017intellectual}
Secundo, G., Perez, S.E., Martinaitis, {\v{Z}}., Leitner, K.H.: An intellectual
  capital framework to measure universities' third mission activities.
\newblock Technological Forecasting and Social Change \textbf{123}, 229--239
  (2017)

\bibitem{spiezia2003measuring}
Spiezia, V.: Measuring regional economies.
\newblock Statistics Directorate of the OECD (2003)

\bibitem{steinmacher2021being}
Steinmacher, I., Balali, S., Trinkenreich, B., Guizani, M., Izquierdo-Cortazar,
  D., Cuevas~Zambrano, G.G., Gerosa, M.A., Sarma, A.: Being a mentor in open
  source projects.
\newblock Journal of Internet Services and Applications \textbf{12}(1), 1--33
  (2021)

\bibitem{su2009spontaneous}
Su, Y.S., Hung, L.C.: Spontaneous vs. policy-driven: The origin and evolution
  of the biotechnology cluster.
\newblock Technological Forecasting and Social Change \textbf{76}(5), 608--619
  (2009)

\bibitem{takhteyev2012coding}
Takhteyev, Y.: Coding places: Software practice in a South American city.
\newblock MIT Press (2012)

\bibitem{takhteyev2010investigating}
Takhteyev, Y., Hilts, A.: Investigating the geography of open source software
  through github.
\newblock Manuscript  (2010)

\bibitem{ITU2019}
{The International Telecommunication Union}: Internet penetration (2019)

\bibitem{toth2021repeated}
T{\'o}th, G., Juh{\'a}sz, S., Elekes, Z., Lengyel, B.: Repeated collaboration
  of inventors across european regions.
\newblock European Planning Studies pp. 1--21 (2021)

\bibitem{trippl2009knowledge}
Trippl, M., T{\"o}dtling, F., Lengauer, L.: Knowledge sourcing beyond buzz and
  pipelines: evidence from the vienna software sector.
\newblock Economic geography \textbf{85}(4), 443--462 (2009)

\bibitem{trujillo2021penumbra}
Trujillo, M.Z., Hébert-Dufresne, L., Bagrow, J.P.: The penumbra of open
  source: projects outside of centralized platforms are longer maintained, more
  academic and more collaborative (2021)

\bibitem{uyarra2016impact}
Uyarra, E., Ramlogan, R.: The impact of cluster policy on innovation.
\newblock In: Handbook of innovation policy impact. Edward Elgar Publishing
  (2016)

\bibitem{valiev2018ecosystem}
Valiev, M., Vasilescu, B., Herbsleb, J.: Ecosystem-level determinants of
  sustained activity in open-source projects: A case study of the pypi
  ecosystem.
\newblock In: Proc. of the 2018 26th ACM Joint Meeting ESEC/FSE (2018)

\bibitem{wachs2021crowdfunding}
Wachs, J., Vedres, B.: Does crowdfunding really foster innovation? evidence
  from the board game industry.
\newblock Technological Forecasting and Social Change \textbf{168} (2021).
\newblock \doi{https://doi.org/10.1016/j.techfore.2021.120747}

\bibitem{wagner2021algo}
Wagner, C., Strohmaier, M., Olteanu, A., Kiciman, E., Contractor, N.,
  Eliassi-Rad, T.: Measuring algorithmically infused societies.
\newblock Nature  (2021)

\bibitem{wanzenbock2020impacts}
Wanzenb{\"o}ck, I., Neul{\"a}ndtner, M., Scherngell, T.: Impacts of eu funded
  r\&d networks on the generation of key enabling technologies: Empirical
  evidence from a regional perspective.
\newblock Papers in Regional Science \textbf{99}(1), 3--24 (2020)

\bibitem{wanzenbock2020framework}
Wanzenb{\"o}ck, I., Wesseling, J.H., Frenken, K., Hekkert, M.P., Weber, K.M.: A
  framework for mission-oriented innovation policy: Alternative pathways
  through the problem--solution space.
\newblock Sci. and Pub. Pol. \textbf{47}(4) (2020)

\bibitem{watney2020clusters}
Watney, C.: Clusters rule everything around me.
\newblock Works in progress (2) (2020)

\bibitem{weterings2006impact}
Weterings, A., Boschma, R.: The impact of geography on the innovative
  productivity of software firms in the netherlands.
\newblock Regional Development in the Knowledge Economy. Routledge, London and
  New York  (2006)

\bibitem{wright2020open}
Wright, N., Nagle, F., Greenstein, S.M.: Open source software and global
  entrepreneurship.
\newblock Harvard Business School Technology \& Operations Mgt. Unit Working
  Paper (20-139), 20--139 (2020)

\bibitem{wu2018all}
Wu, L., Jin, F., Hitt, L.M.: Are all spillovers created equal? a network
  perspective on information technology labor movements.
\newblock Management Science \textbf{64}(7), 3168--3186 (2018)

\end{thebibliography}

\end{document}